\documentclass[reprint,aps,prb,superscriptaddress,english,nolongbibliography]{revtex4-2}

\usepackage{graphicx}
\usepackage{amsmath,amssymb,bm,mathrsfs,amsfonts}
\usepackage{natbib}
\usepackage[colorlinks=true,citecolor=blue,urlcolor=blue,linkcolor=blue]{hyperref}

\usepackage{color,colortbl,hhline}
\usepackage[table,xcdraw,dvipsnames]{xcolor}
\usepackage[caption=false,subrefformat=parens,labelformat=empty]{subfig}

\usepackage[normalem]{ulem}

\usepackage[utf8]{inputenc}
\usepackage{fontenc}

\newcommand{\sbra}[1]{\langle #1 |}
\newcommand{\sket}[1]{| #1 \rangle}

\newcommand{\bra}[1]{\left\langle #1 \right|}
\newcommand{\ket}[1]{\left| #1 \right\rangle}

\newcommand{\av}[1]{\left \langle #1 \right \rangle}

\usepackage{tikz}

\newcommand{\urmove}[0]{%
\begin{tikzpicture}[baseline=0.1ex]%
\draw[thick, red] (0,1.5ex) -- (1.5ex,1.5ex);%
\draw[thick, red] (1.5ex,1.5ex) -- (1.5ex,0);%
\end{tikzpicture}%
}

\newcommand{\ulmove}[0]{%
\begin{tikzpicture}[baseline=0.1ex]%
\draw[thick, red] (0,1.5ex) -- (1.5ex,1.5ex);%
\draw[thick, red] (0,1.5ex) -- (0,0);%
\end{tikzpicture}%
}

\newcommand{\dlmove}[0]{%
\begin{tikzpicture}[baseline=0.1ex]%
\draw[thick, red] (0,1.5ex) -- (0,0);%
\draw[thick, red] (0,0) -- (1.5ex,0);%
\end{tikzpicture}%
}

\newcommand{\drmove}[0]{%
\begin{tikzpicture}[baseline=0.1ex]%
\draw[thick, red] (0,0) -- (1.5ex,0);%
\draw[thick, red] (1.5ex,1.5ex) -- (1.5ex,0ex);%
\end{tikzpicture}%
}

\newcommand{\vvmove}[0]{%
\begin{tikzpicture}[baseline=0.1ex]%
\draw[thick, red] (0,0) -- (0,1.5ex);%
\draw[thick, red] (1.5ex,0) -- (1.5ex,1.5ex);%
\end{tikzpicture}%
}

\newcommand{\hhmove}[0]{%
\begin{tikzpicture}[baseline=0.1ex]%
\draw[thick, red] (0,0) -- (1.5ex,0);%
\draw[thick, red] (1.5ex,1.5ex) -- (0,1.5ex);%
\end{tikzpicture}%
}

\begin{document}

\title{Slow melting of a disordered quantum crystal}

\author{Federico Balducci}
\email{fbalducc@sissa.it}
\affiliation{SISSA -- via Bonomea 265, 34136, Trieste, Italy}
\affiliation{INFN Sezione di Trieste -- Via Valerio 2, 34127 Trieste, Italy}
\affiliation{The Abdus Salam ICTP -- Strada Costiera 11, 34151, Trieste, Italy}
\author{Antonello Scardicchio}
\affiliation{INFN Sezione di Trieste -- Via Valerio 2, 34127 Trieste, Italy}
\affiliation{The Abdus Salam ICTP -- Strada Costiera 11, 34151, Trieste, Italy}
\author{Carlo Vanoni}
\affiliation{SISSA -- via Bonomea 265, 34136, Trieste, Italy}
\affiliation{INFN Sezione di Trieste -- Via Valerio 2, 34127 Trieste, Italy}
\affiliation{The Abdus Salam ICTP -- Strada Costiera 11, 34151, Trieste, Italy}

\date{\today}

\begin{abstract}
    The melting of the corner of a crystal is a classical, real-world, non-equilibrium statistical mechanics problem which has shown several connections with other branches of physics and mathematics. For a perfect, classical crystal in two and three dimensions the solution is known: the crystal melts reaching a certain asymptotic shape, which keeps expanding ballistically. In this paper, we move onto the quantum realm and show that the presence of quenched disorder slows down severely the melting process. Nevertheless, we show that there is no many-body localization transition, which could impede the crystal to be completely eroded. We prove such claim both by a perturbative argument, using the forward approximation, and via numerical simulations. At the same time we show how, despite the lack of localization, the erosion dynamics is slowed from ballistic to logarithmic, therefore pushing the complete melting of the crystal to extremely long timescales.
\end{abstract}

\maketitle

\section{Introduction}

Research on quantum dynamics of extended objects, in particular in situations far from equilibrium, has seen a surge in activity in the last years, pushed by two, apparently independent advances. On one side, the experimental techniques to control mesoscopic systems have improved significantly in the fields of cold atoms~\cite{Bloch2008Many,lewenstein2012ultracold}, trapped ions~\cite{Leibfried2003Quantum,Browaeys2020Many}, nitrogen-vacancy (NV) centers in diamond~\cite{BarGill2013Solid,schirhagl2014nitrogen,Hernandez2020quantum}, superconducting qubits~\cite{devoret2004superconducting}, and others~\cite{Noh2016Quantum}. On the other side, novel theoretical results have been obtained regarding the dynamics of strongly interacting quantum systems, both in presence and in absence of quenched disorder. In the first case, the work~\cite{basko2006metal} made a strong case for the breakdown of ergodicity in presence of disorder, even if interactions are present. This many-body localized (MBL) phase~\cite{huse2015review,abanin2019colloquium} is in stark contrast with the previously assumed picture, for which a generic interacting system would be ergodic in the sense of the eigenstate thermalization hypothesis (ETH)~\cite{deutsch1991quantum,srednicki1994chaos}. The same phenomenology has been further observed even in models \emph{without} quenched disorder, but presenting e.g.\ kinetic constraints~\cite{Smith2017Disorder,Brenes2018Many,Pancotti2020Quantum}, Hilbert space fragmentation~\cite{Nandkishore2019Fractons,Sala2020Erdodicity}, or large energy gaps throughout the spectrum~\cite{vanNieuwenburg2019From,Schulz2019Stark}.

The picture emerging from these recent advances is that isolated quantum systems, out of equilibrium and (possibly) in the presence of disorder, can display a serious suppression of transport. Generically, they can pass from the usual diffusive dynamics at small disorder, to subdiffusive transport~\cite{agarwal2015anomalous,znidaric2016diffusive,luitz2017ergodic,taylor2021subdiffusion}, and then finally to a localized regime~\cite{anderson1958absence}, becoming effectively integrable systems~\cite{serbyn2013local,huse2014phenomenology,ros2015integrals,imbrie2017local}. So far, evidence for the MBL phase beyond perturbation theory~\cite{basko2006metal,ros2015integrals} consists in many numerical results on small, one dimensional spin chains~\cite{znidaric2008many,pal2010many,de2013ergodicity,kjall2014many,luitz2015many}, some experimental results for larger systems but smaller times~\cite{Schreiber2015Observation,Choi2016Exploring,Smith2016Many}, and a single analytic work~\cite{imbrie2016many,imbrie2016diagonalization} proving MBL in a spin chain under some reasonable assumptions. 

While the situation is slowly getting under control in one dimension (but with many caveats and uncertainties~\cite{Suntajs2020Quantum,Suntajs2020Ergodicity,panda2020can,Abanin2021Distinguishing,Crowley2020Constructive,Morningstar2022Avalanches,Sierant2022Challenges}), in dimensions two and higher, the existence of MBL beyond the original paper~\cite{basko2006metal} and a few others~\cite{chandran2016many,Nandkishore2017Many,Gopalakrishnan2019Instability,Gopalakrishnan2020Dynamics,Artiaco2021Signatures} is even more questionable. The main issue is that in $2d$ numerical results are limited to extremely small systems~\cite{Wahl2019Signatures}, and at the same time non-perturbative effects are thought to be stronger~\cite{de2017many,de2017stability,Thiery2018many}. Some counterexamples are provided by two studies of dimer models in $2d$~\cite{theveniaut2020transition,pietracaprina2021probing} which, because of the slow (though still exponential) growth of their Hilbert space dimension, can be studied up to 100 spins or so; and by some recent investigations on quasi-periodically-modulated $2d$ models~\cite{Agrawal2022Note,Strkalj2022Coexistence,Crowley2022Mean}. Such studies show an MBL transition with the same confidence that it is seen in spin chains, giving hope that disorder can indeed localize in $2d$ following the same route that works in $1d$, at least for some particular microscopic Hamiltonians. 

In this work, we consider under the perspective of MBL the process of melting of the corner of an imperfect, two-dimensional quantum crystal. The process of melting of \emph{classical} crystals is a wide-studied phenomenon in statistical mechanics~\cite{bray1994theory,onuki2002phase} and mathematical physics~\cite{Cerf2001Low,Kenyon2006Dimers}, with connections ranging from the theory of random integer partitions~\cite{Okounkov2000Random,Okounkov2001Infinite,Okounkov2003Uses} to determinantal point processes~\cite{okounkov2003correlation,ferrari2003step,Spohn2006Exact,torquato2008point,scardicchio2009statistical}, and even to Calabi-Yau manifolds~\cite{Okounkov2006Quantum,Ooguri2009Crystal}. Alongside previous works~\cite{Balducci2022Localization,Balducci2022Interface}, we model the melting process by the strong-coupling limit of the $2d$ quantum Ising model in both transverse and longitudinal magnetic fields. In particular, via a Schrieffer-Wolff transformation~\cite{Schrieffer1966Relation} one can obtain an effective Hamiltonian, that is particularly suitable for interpreting the process in terms of the motion of the ``crystal-liquid'' interface. Such Hamiltonian is in the same family of constrained PXP models~\cite{Fendley2004Competing,Lesanovsky2012Liquid,Surace2021Exact}, arising in the context of ultracold Rydberg atoms~\cite{Lukin2001Dipole}. In the presence of disorder, PXP models show resilience towards localization already in $1d$~\cite{sierant2021constraints}. The explanation relies on the fact that the local disorder \emph{before} the constraints are applied maps to \emph{non-local} terms in the Hamiltonian, which escape in this way the usual arguments leading to localization in the perturbative limit. This is a first clue that makes us suspect that crystal melting cannot be stopped by disorder, no matter how strong the latter can be. In this paper, we will exactly prove this working hypothesis: the dynamics of the crystal melting gets only slowed down---albeit quite dramatically---never stopping at any finite value of disorder.

To prove our claim, we proceed as follows. After having introduced the model, which is the quantum Ising model in two spatial dimensions, we consider the evolution of a particular type of initial condition, under the approximation that the Ising coupling $J$ be the largest scale in the problem, the other two being the longitudinal ($h$) and transverse ($g$) magnetic fields. Within this approximation, the states of the Hilbert space can be put in one-to-one correspondence with Young diagrams, thereby reducing considerably the growth of the Krylov subspaces for the evolution. This allows us to go to relatively large system sizes, and explore the dynamical and eigenstate properties as the amount of disorder is increased. We find that, for any given system size, the eigenstate properties show some signs of localization, at least for sufficiently large disorder. However, the disorder strength for which localization is seen grows with the system size in a way which seems to indicate that no transition to MBL is present in the thermodynamic limit. Because of this, the delocalized phase emerging in such limit is rather peculiar in nature, as the dynamics is extremely slow: for example, the expected number of spin flips at time $t$ grows like $\sim \ln (gt)$, irrespective of the value of disorder. This is in contrast with other situations in which the delocalized side shows transport dictated by continuously changing exponents, that are functions of the disorder strength (see for example Refs.~\cite{luitz2017ergodic,panda2020can,taylor2021subdiffusion}). We support these numerical findings with an analytical argument, employing the forward approximation in the locator expansion of the resolvent.

Our findings are relevant for several reasons. First, they show that when the process of melting of a crystal is quantum-coherent, then it cannot be stopped even by the presence of arbitrarily strong quenched disorder. Second, our work shows that generic PXP models do not likely present any stable MBL phase in two dimensions, despite this feature is very difficult to be inferred from the dynamical evolution alone. Indeed, the delocalized phase suffers of a severe dynamical slowdown, which could be easily misinterpreted for localization if considered alone (e.g.\ in an experimental setting, where eigenstate properties are difficult to access). Third, our findings hint at the conclusion that quenched disorder \emph{and} dynamical constraints, when combined, prevent the occurrence of a stable MBL phase in two dimensions. While our results do not constitute a real proof of this latter statement, they nevertheless provide solid evidence. In this respect, our work is one of the very few numerical studies (others that we are aware of are Refs.~\cite{theveniaut2020transition,pietracaprina2021probing}) that is able to explore two-dimensional system sizes, which are not too small to draw any possible conclusion on the thermodynamic limit. 

The rest of the paper is organized as follows. In Sec.~\ref{sec:model2} we introduce the model and in Sec.~\ref{sec:PXP} we specialize it to the strong coupling limit. In Sec.~\ref{sec:Young_diagrams} we discuss the connection between the dynamics of a corner interface and the growth of Young diagrams, while is Sec.~\ref{sec:fermions} we show how to describe the corner evolution using a $1d$ fermionic chain. Subsequently, in Sec.~\ref{sec:perturbation_theory} we move to discuss the forward approximation for the model under consideration, presenting in Sec.~\ref{sec:FA} the analytic treatment and in Sec.~\ref{sec:FA_results} the comparison with numerical results. Then, we move to a detailed presentation of the numerical results both for the spectrum, in Sec.~\ref{sec:spectrum_numerics}, and for the dynamics, in Sec.~\ref{sec:dynamics_numerics}. Finally, in Sec.~\ref{sec:discussion} we discuss the limits of validity of the results presented when the strong coupling limit is relaxed (Sec.~\ref{sec:approximations}) and the comparison with corresponding classical models (Sec.~\ref{sec:corner_growth_models}), giving some final considerations in Sec.~\ref{sec:conclusions}.

\section{Model}
\label{sec:model}

\subsection{A disordered quantum solid undergoing melting}
\label{sec:model2}

\begin{figure}
    \centering
    \includegraphics[width=0.5\textwidth]{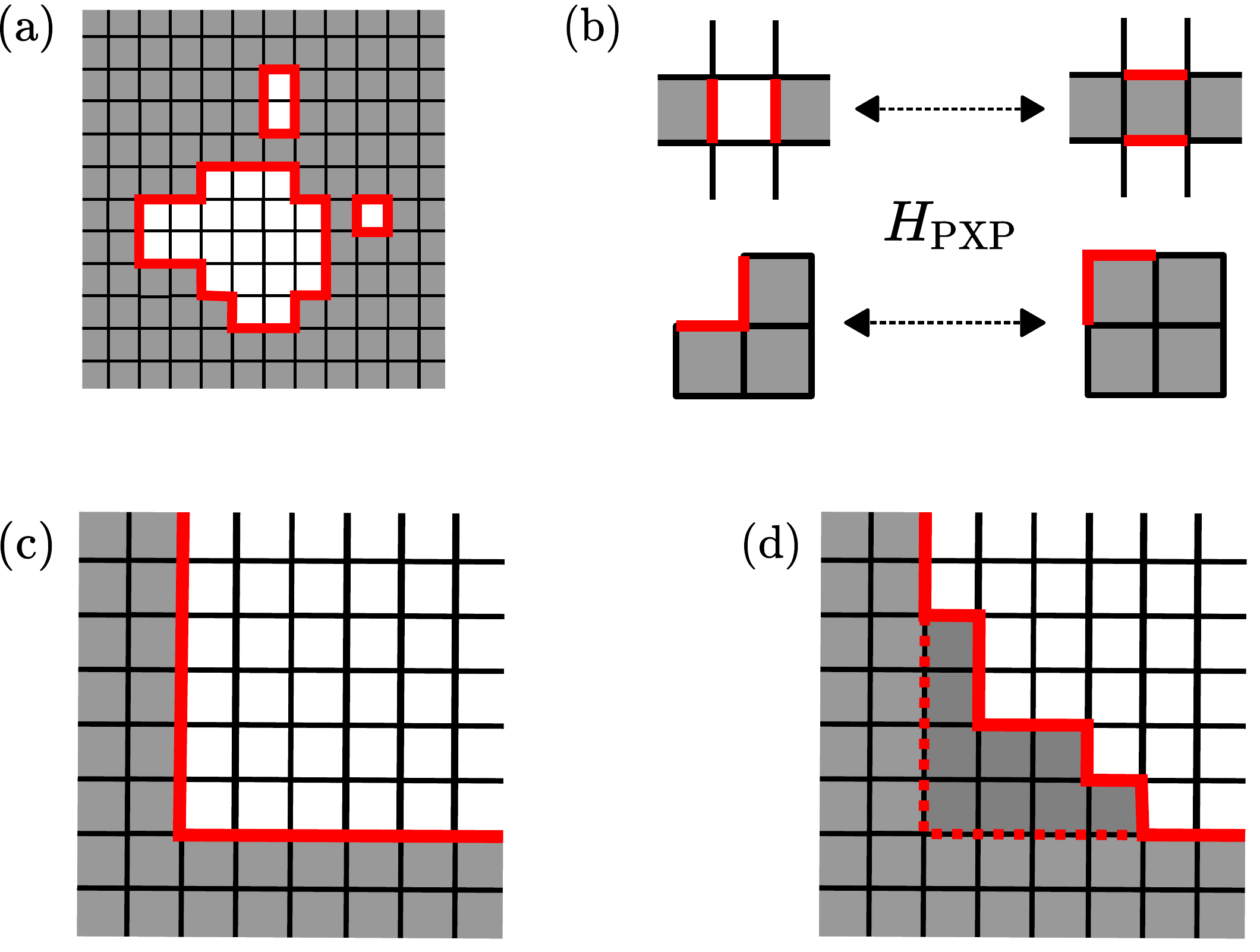}
    \caption{(a) Generic configuration diagonal in the $\sigma^z$ basis, made of $3$ disconnected bubbles of ``down'' spins (in white) surrounded by ``up'' spins (in gray). (b) Visual representation of the hopping terms of the Hamiltonian Eq.~\eqref{eq:H_PXP}. The top row represents moves that break a bubble into two pieces (or join them), and will not be considered in our discussions. The bottom row represents moves that give dynamics to the corner, allowing it to melt: we will focus our attention on these ones. (c) The initial configuration we will consider: an infinite, right-angled wedge. (d) One of the possible configurations reached from the wedge in the melting process.}
    \label{fig:PXP}
\end{figure}

As anticipated in the Introduction, we are interested in the dynamics of melting of a two-dimensional, disordered quantum crystal. As done commonly in the literature~\cite{Cerf2001Low,Kenyon2006Dimers,Okounkov2001Infinite,Spohn2006Exact}, we consider the melting process starting from the tip of an infinite, right-angled wedge, see Fig.~\ref{fig:PXP}. More general finite- and infinite-size initial configurations could be treated with similar tools, see for more details Refs.~\cite{Balducci2022Localization,Balducci2022Interface}. We describe the solid, non-melted part of the crystal via ``up'' Ising spins $\sigma^z_i = +1$, $i\in \mathbb{Z}^2$, and the melted part via ``down'' spins $\sigma^z_i = -1$ ($\sigma^{x,y,z}$ are Pauli matrices). The Hamiltonian is that of the two-dimensional Ising model on a square lattice, with a constant transverse field $g$ and a random longitudinal field $h_i$:
\begin{equation}
    \label{eq:H_Ising}
    H_{\mathrm{Ising}} = -J\sum_{\langle ij \rangle} \sigma^z_i \sigma^z_j + \sum_i h_i \sigma^z_i + g\sum_i \sigma^x_i.
\end{equation}
The physical interpretation of the terms appearing in the Hamiltonian is rather straightforward. The $g$ term lets the spins flip freely between $+1$ and $-1$, thus neither the melted phase nor the crystal is preferred at this level (we stress that we want to describe the \emph{quantum-coherent} process of melting \emph{in real time}, thus we need time reversibility). The presence of the ferromagnetic coupling term ($J>0$ throughout the paper), however, favours the formation of bubbles of aligned spins, contrasting to a certain extent the action of the $g$ flips. Finally, we introduce disorder in the form of a random longitudinal field $h_i \in [-W/2,W/2]$, with a uniform distribution: this models the presence of impurities by assigning a different energy cost for the addition/removal of a particle at each site $i \in \mathbb{Z}^2$. Notice that our choice of the Hamiltonian~\eqref{eq:H_Ising} implies that particles in the liquid state \emph{do not lose phase coherence}, as they are represented, in a rather simplistic way, by immobile ``down'' spins, that do not wander around and interact with one another. This choice is made so to boost the quantum coherence of the model, which otherwise should be described as an open quantum system.

\subsection{Strong-coupling limit and effective PXP description}
\label{sec:PXP}

Throughout the paper, we will assume that the strong-coupling limit $J \gg g \sim W$ hold. This assumption is necessary to make sense of a \emph{quantum-fluctuating interface}, that clearly separates the solid and melted phases; otherwise, one could not speak of a melting process altogether. 

When the coupling $J$ is very strong, the Hilbert space of the model effectively decomposes into sectors identified by the length of the domain walls, i.e.\ the number of violated Ising bonds~\cite{Yoshinaga2021Emergence,Hart2022Hilbert,Balducci2022Localization,Balducci2022Interface}. More precisely, one can introduce the operator $\mathcal{L} = \sum_{\langle i,j \rangle} (1 - \sigma^z_i \sigma^z_j )/2$, which exactly measures the number of frustrated Ising bonds. The operator $\mathcal{L}$ is the combined length of the strings/domain walls, and it is a conserved quantity in the limit $J \to \infty$. Indeed, $\mathcal{L}$ is very closely related to the interaction energy in the original model, and by unitarity it must be conserved: the excess energy cannot be compensated by other means.

When instead $J$ is large but finite, $\mathcal{L}$ is only approximately conserved. However, while for the ordered case with field $h$ this is a singular limit (since the energy of a string of length $\ell$ is $J\ell$ and the volume energy contribution $\sim h \ell^2$ is always dominant), for our disordered model with average field $\av{h}=0$, the volume energy contribution is typically of order $\av{h^2}^{1/2}\ell$. Thus, in the limit $J\gg h$ the string length can be conserved to high accuracy. 

As just described, through the operator $\mathcal{L}$ the Hilbert space is split into sectors of equal domain-wall length and, if the initial condition is supported within only one of those sectors, the dynamics will be confined in it for all times. In Refs.~\cite{Yoshinaga2021Emergence,Hart2022Hilbert} the fragmentation of the Hilbert space of the clean version of the model~\eqref{eq:H_Ising} was studied in great detail, showing in particular that the \emph{Krylov subspaces} represent an even finer scale wrt.\ the eigenspaces of $\mathcal{L}$. In Refs.~\cite{Balducci2022Localization,Balducci2022Interface}, instead, the \emph{dynamical implications} of the fragmentation were inspected: here, we will build on these latter works, showing how the picture is modified by the introduction of disorder.

The Hilbert space fragmentation in the strong-coupling limit can be formally described through a Schrieffer-Wolff (SW) transformation~\cite{Schrieffer1966Relation}, that accounts for the Ising interaction in a perturbative way. In terms of the operator $\mathcal{L}$, the Hamiltonian $H_{\mathrm{eff}}$ generated by the SW transformation, order by order in $J^{-1}$, will be such that $[H_{\mathrm{eff}},\mathcal{L}]=0$. To lowest order, the effective Hamiltonian one finds in any Krylov subspace is of the PXP form~\cite{Balducci2022Localization,Balducci2022Interface}: 
\begin{multline}
    \label{eq:H_PXP}
    H_{\mathrm{PXP}} = \sum_i h_i \sigma_i^z \\ 
    + g \sum_i \Big( \sket{\ulmove} \sbra{\drmove} + \sket{\dlmove} \sbra{\urmove} + \sket{\hhmove} \sbra{\vvmove} + \mathrm{h.c.} \Big).
\end{multline}
Above, we have introduced a convenient graphical notation to indicate spin flips. Indeed, one can easily get convinced that spin flips can take place only next to a up/down spin border, as in Fig.~\hyperref[fig:PXP]{\ref{fig:PXP}a}, and are only of the form indicated by the shapes in Eq.~\eqref{eq:H_PXP} and Fig.~\hyperref[fig:PXP]{\ref{fig:PXP}b}.

The next order in the SW transformation encompasses the first $O(1/J)$ corrections. However, the resulting SW Hamiltonian is rather complicated, and probably of little practical use in general situations. In Refs.~\cite{Balducci2022Localization,Balducci2022Interface} such Hamiltonian is derived, but only for a class of physically relevant Krylov sectors (which comprise the one investigated in this work). While we could study also in this paper the effects of a finite $J$, we believe that such effects would entail just a \emph{quantitative} modification of the results presented, while leaving the physical picture unchanged. Therefore, in the following we will always neglect the $O(1/J)$ corrections, while leaving to Sec.~\ref{sec:approximations} a brief informal discussion of their possible implications.

So far, we have argued that one can pass from the Hamiltonian of the full $2d$ quantum Ising model, Eq.~\eqref{eq:H_Ising}, to the effective Hamiltonian, Eq.~\eqref{eq:H_PXP}, capturing the dynamics of domain walls in the original model, when the strong coupling limit is considered.
Before moving on, let us remind that PXP Hamiltonians in $1d$ have shown some form of slow dynamics in either numerics or experiments~\cite{bernien2017probing,turner2018quantum,moudgalya2018exact,ho2019periodic,khemani2019signatures}, and presence of ``scars'' in the spectrum, i.e.\ atypical eigenstates (e.g.\ with atypically low entanglement entropy). At the same time, for PXP models both the spectrum as a whole, and the dynamics at finite energy density, are ergodic. Such ergodicity is resistant also to the introduction of quenched disorder~\cite{sierant2021constraints}: this is a consequence of the fact that the disorder maps, in an \emph{unconstrained} basis of states, to generic, non-local interaction terms. This feature will be present also in the $2d$ model under consideration, thus we refer to Sec.~\ref{sec:Young_and_fermions} for a detailed discussion.

\section{Mapping to Young diagrams and to lattice fermions}
\label{sec:Young_and_fermions}

As stated in the Introduction, we are interested in the dynamics of melting generated by the Hamiltonian \eqref{eq:H_Ising} (or equivalently Eq.~\eqref{eq:H_PXP}), starting from a particular type of initial condition: a corner made of ``up'' spins, in a sea of otherwise ``down'' spins, see Fig.~\hyperref[fig:PXP]{\ref{fig:PXP}c}. This configuration is physically relevant, as it is one of the simplest crystal shapes whose melting can be studied. In this Section, we discuss two mappings of the Krylov subspace containing such corner-shaped interface: a mapping to Young diagrams in Sec.~\ref{sec:Young_diagrams}, and one to lattice fermions in Sec.~\ref{sec:fermions}. The usefulness of such mappings will become clear as we proceed.

\subsection{Young diagrams}
\label{sec:Young_diagrams}

The fragmentation into Krylov subspaces, briefly outlined in the previous Section, represents a huge source of simplification for the full, $2d$ problem. A particularly neat example is given by the initial state whose evolution we aim at describing, viz.\ a right-angled, infinite corner (see Fig.~\hyperref[fig:PXP]{\ref{fig:PXP}c}). For this case in particular, only the moves $(\sket{\ulmove} \sbra{\drmove} + \mathrm{h.c.})$ and $(\sket{\dlmove} \sbra{\urmove} + \mathrm{h.c.})$ in Eq.~\eqref{eq:H_PXP} are allowed, and all the states in the Krylov subspace are in one-to-one correspondence with \emph{Young diagrams}. Let us recall that, by definition, a Young diagram is a collection of boxes, arranged in left-justified rows, and stacked in non-increasing order of length~\cite{Fulton1996Young}. The mapping to Young diagrams is quite transparent; a detailed explanation can be found in Refs.~\cite{Balducci2022Localization,Balducci2022Interface}. We also recall \emph{en passant} that the Young diagrams of size $N$ are in one-to-one correspondence with the integer partitions of $N$.

\begin{figure}
    \centering
    \includegraphics[width=\columnwidth]{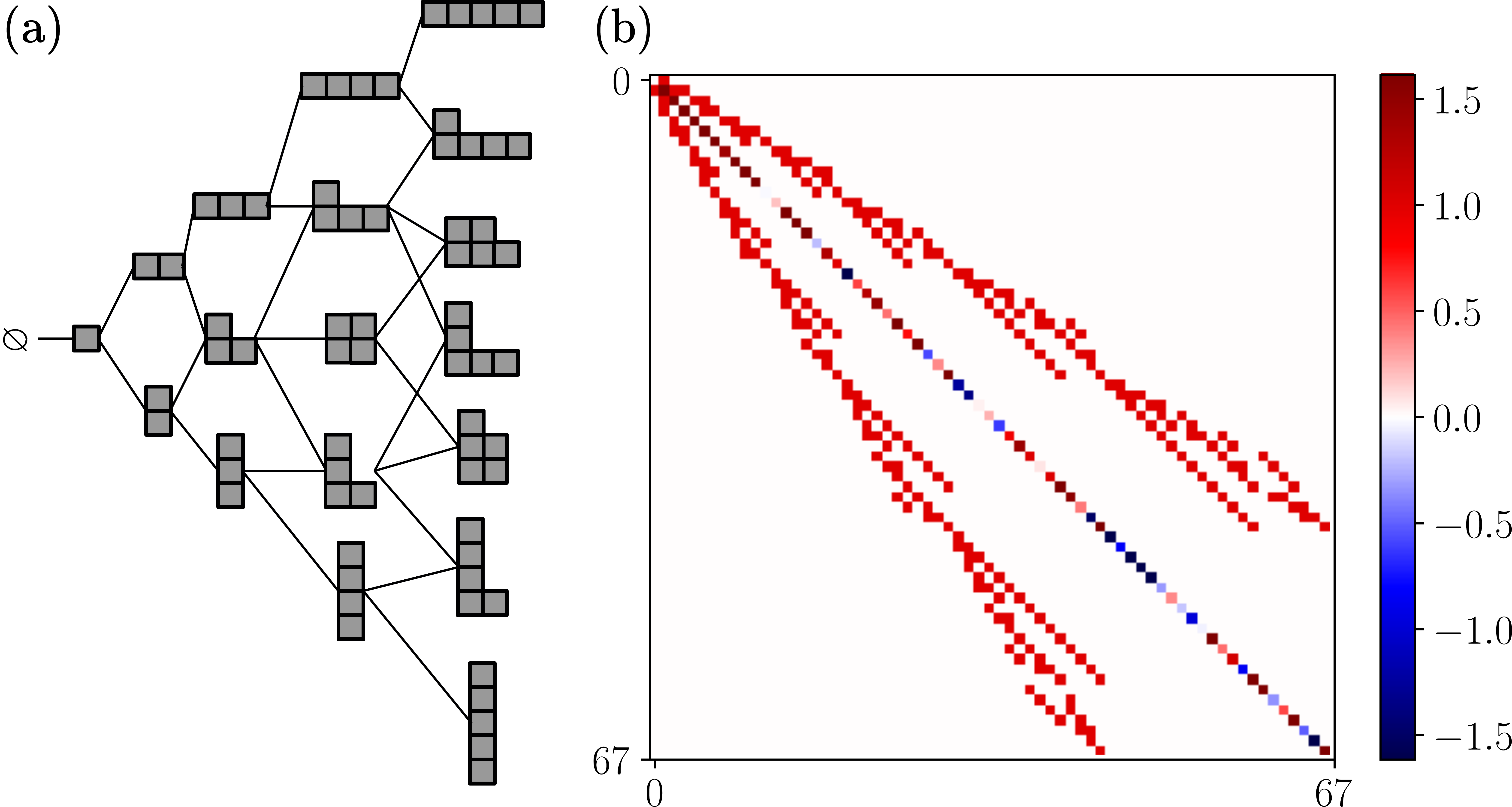}
    \caption{(a) Young lattice, i.e.\ the set of Young diagrams where two of them are connected if differing by a single box. In figure the lattice up to $N=5$ is represented.
    (b) Matrix plot of the Hamiltonian Eq.~\eqref{eq:YD_Hamiltonian} up to $N=8$, corresponding to a Hilbert space of dimension 67. The off-diagonal elements correspond to the adjacency matrix of the Young lattice, and are all set to $g\equiv 1$, while the diagonal part is determined by the disordered magnetic field as detailed in Sec.~\ref{sec:Young_diagrams}.}
    \label{fig:YoungLattice}
\end{figure}

Thanks to the mapping, the quantum dynamics which makes the crystal wedge melt can be described equivalently by the hopping on the space of Young diagrams $\mathcal{D}$, see Fig.~\ref{fig:YoungLattice}. The initial state, viz.\ the full wedge, is the empty Young diagram $D = \emptyset$. Then, the energy of a diagram $D \in \mathcal{D}$ is given by the sum of the longitudinal fields on the ``blocks'' composing the diagram:
\begin{equation}
    \label{eq:E_D}
    E_{D} = \sum_{i \in D} h_i.
\end{equation}
The rate of hopping between two Young diagrams $D,D'$ is $g$ if they are connected by a single block addition or deletion (neighbouring diagrams), or zero otherwise. Therefore, the adjacency matrix of the Young lattice has non-zero elements only between the set of diagrams of size $N$, call it $\mathcal{D}_{N}$, and that of size $N-1$ ($\mathcal{D}_{N-1}$) or $N+1$ ($\mathcal{D}_{N+1}$); see Fig.~\hyperref[fig:YoungLattice]{\ref{fig:YoungLattice}b} for a sketch.

In the end, one is left with a Hamiltonian operator, acting on the Hilbert space $\mathcal{H}_\mathcal{D}$ built on the set of diagrams $\mathcal{D}$, i.e.\ the Krylov subsector of the original Ising model that contains the infinite wedge:
\begin{equation}
    \label{eq:YD_Hamiltonian}
    H_{\mathcal{D}} = g \sum_{\langle D,D' \rangle} \sket{D} \sbra{D'} + \sum_D E_D \sket{D} \sbra{D}.
\end{equation}
The net gain is that the dimension of $\mathcal{H}_{\mathcal{D}}$ is much smaller than that of the full Hilbert space of all the spins configurations $\{ \sigma_i \}$ on the plane. Indeed, let us denote the dimension of the Hilbert subspace, made of diagrams composed of exactly $N$ squares, as $d_N := \dim \mathcal{H}_{\mathcal{D}_N}$. It follows that, for the diagrams made up at most of $N$ squares, one has to compute the cumulative $\bar{d}_N := \sum_{k=0}^N d_k$. Thus, from the Hardy-Ramanujan asymptotic formula for partitions, one finds $\bar{d}_N \simeq \exp{(\pi \sqrt{2N/3})} / \sqrt{8 \pi^2 N}$: the mild, stretched-exponential growth of such numbers will enable us to reach system sizes of up to $N = 36$ spins. Notice that such dimensions correspond to a \emph{vanishing entropy density} in the original model, since $s = \ln(\bar{d}_N)/N \sim N^{-1/2}$. In other constrained models (including the $2d$ dimer models of Refs.~\cite{theveniaut2020transition,pietracaprina2021probing}) the growth of Krylov sectors is instead exponential, with a finite entropy density.

Before moving on, let us remark that the approach outlined above, i.e.\ passing from the original interacting model to an hopping problem on the Hilbert space graph, is a common practice in the field of many-body physics, and in particular of MBL~\cite{pietracaprina2016forward}. For more standard quantum spin chains with particle number conservation, one usually restricts to the half-filling sector, thus obtaining the subset of the hypercube with an equal number of positive and negative vertices as graph---eventually with a disordered, correlated chemical potential if the original model is disordered itself. In the case under consideration, instead, the graph obtained is the Young diagrams lattice of Fig.~\ref{fig:YoungLattice}, another subset of the hypercube but with very different connectivity properties wrt.\ the former: this can be already guessed from the scaling with $N$ of the number of vertices and edges~\cite{Griffiths2011Hook}. Let us also mention that this hopping problem is very different from XXZ-type models \emph{on random graphs themselves}~\cite{Cuevas2012Level}, since one has already got rid of interactions by passing to the graph.

\begin{figure}
    \centering
    \includegraphics[width=\columnwidth]{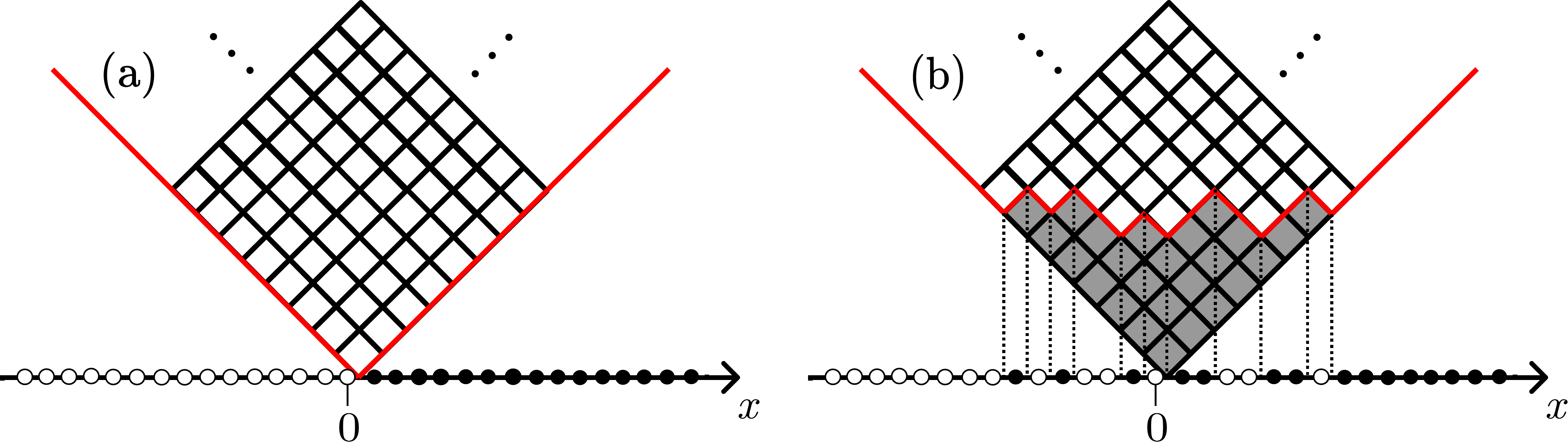}
    \caption{Mapping of $2d$ configurations dynamically connected to the corner, and thus of Young diagrams, to fermions. 
    (a) The initial configuration, i.e. an infinite wedge, is mapped to the domain wall state $\ket{\cdots 0 0 0 1 1 1 \cdots}$ on the chain. 
    (b) A generic state (in this case, the Young diagram corresponding to the partition $\{7,6,4,3,3,1,1\}$) is mapped to a fermion configuration according to the procedure described in the main text.}
    \label{fig:Mapping}
\end{figure}

\subsection{Lattice fermions}
\label{sec:fermions}

It is a classic result of combinatorics that Young diagrams can be mapped to a spin-$1/2$ chain~\cite{Rost1981Non} or, equivalently, to a fermionic chain via a Jordan-Wigner transformation, see also Fig.~\ref{fig:Mapping}. This has allowed for an analytic solution for the limiting shape of the crystal wedge (for a clean system)~\cite{Balducci2022Localization,Balducci2022Interface}, and unveiled connections to the mathematics of random integer partitions~\cite{Okounkov2000Random,Okounkov2003Uses,Okounkov2006Quantum}. In the case of a clean system, in particular, the Hamiltonian \eqref{eq:H_Ising} maps to free fermions on the chain. In the disordered case we are now studying, instead, the integrability will be lost but the mapping, which is a form of \emph{holography} between a $2d$ problem and a $1d$ problem, retains its usefulness in simplifying the description of the problem, both for a numerical and an analytical treatment. Therefore, we will briefly describe it here. 

Let us start with the null Young diagram $D=\emptyset$: as shown in Fig.~\hyperref[fig:Mapping]{\ref{fig:Mapping}a}, it is associated to a domain wall centered in 0 on the chain. Then, by moving on the line particles to the left, or holes to the right, to each and every $1d$ fermion configuration at half filling there corresponds a Young diagram, as in Fig.~\hyperref[fig:Mapping]{\ref{fig:Mapping}b}. 

For what concerns the Hamiltonian, the hopping term becomes associated to simple nearest-neighbour hoppings on the chain (whose fermionic operators we represent as $\psi_x$, $x \in \mathbb{Z}$)
\begin{equation}
    g \sum_{\langle D,D' \rangle} \sket{D} \sbra{D'}
    \quad \longleftrightarrow \quad 
    g \sum_{x} \left( \psi_x^\dagger \psi_{x+1} + \mathrm{h.c.} \right) .
\end{equation}
The energy $E_D\ket{D}\bra{D}$, on the other hand, has no simple interpretation as a local term. Instead, it is a generic operator which involves all the fermions, through their number operator $n_x = \psi_x^\dagger \psi_x$:
\begin{equation}
    \label{eq:interactions}
    E_D \sket{D} \sbra{D} 
    \quad \longleftrightarrow \quad 
    E(n_x,n_y, \dots)
\end{equation}
where $x,y,...$ are the indices of the sites ``touched'' by the diagram $D$. This non-locality of the disordered potential terms, already anticipated in the Introduction, is typical of PXP models~\cite{sierant2021constraints}, and it comes from the interplay of dynamical constraints and local fluctuations in the potential energy. In one spatial dimension, it was proven to be the cause of the absence of a MBL phase~\cite{sierant2021constraints}: indeed, the presence of non-local interactions on the chain makes the model evade all the arguments in favor of ergodicity breaking. We believe that the same happens in our $2d$ setting, since the perturbative arguments supporting MBL work equally in any dimension, while the non-perturbative effects that destabilize MBL are stronger.

In view of the above, it is quite surprising to remark that, on the contrary, in the clean case $h_i \equiv h$ the mapping simplifies to
\begin{equation}
    \sum_D E_D \sket{D} \sbra{D} 
    \quad \longleftrightarrow \quad 
    -2 h \sum_x x \, \psi_x^{\dagger} \psi_x^{\phantom{\dagger}}.
\end{equation}
Therefore, for a uniform field $h \neq 0$ the melting dynamics will be Stark-localized, as found in Refs.~\cite{Balducci2022Localization,Balducci2022Interface}. Moreover, for $h \gtrsim 1$ the finite-$J$ corrections are likely incapable of thermalizing the system, which therefore enters a Stark-MBL phase~\cite{Balducci2022Localization,Balducci2022Interface}. We see therefore that the presence of disorder is \emph{assisting} the thermalization, since it breaks the integrability (in the sense of free fermions) of the model, while it is not able to make the model athermal by itself, due to its non-local nature.

To conclude this Section, we remark that the mapping of the $2d$ dynamics onto a line of fermions is interesting for several reasons. First, as said above it constitutes a great simplification of the problem, as it enables a $1d$ effective description, amenable of much more analytical control. Second, the $2d$ dynamics induces on the fermions a rather particular type of dynamics, interesting by itself, which we set up to investigate in the next Sections. Third, as remarked also in Refs.~\cite{Balducci2022Localization,Balducci2022Interface}, the mapping is a form of \emph{holography}~\cite{Maldacena1999Large}, which surely deserves a better investigation, in view of the intense interest of the last years on such phenomena, especially in presence of quenched disorder~\cite{Sachdev_1993,Kitaev2015Holography,Arean2016Holographic}.

\section{Perturbation theory estimates}
\label{sec:perturbation_theory}

It is becoming clear, as the discussion unfolds, that the melting of an infinite quantum crystal wedge does not undergo a localization phenomenon, even if it may be severely slowed down by disorder. Therefore, as a first thing we perform a perturbative estimate for the critical disorder strength $W_c$ of a putative MBL transition, showing that such $W_c$ flows to infinity as the thermodynamic limit is approached. To do so, we employ the so-called \emph{forward approximation} (FA)~\cite{anderson1958absence,abou1973self,pietracaprina2016forward,Scardicchio2017Perturbation}, which consists in calculating the Green's functions to lowest order in the hopping among localized orbitals. For the sake of being self-contained, we review briefly the main ideas of the FA in Sec.~\ref{sec:FA}, and then discuss the implications for our system in Sec.~\ref{sec:FA_results}.

\subsection{Brief description of the forward approximation}
\label{sec:FA}

In the FA, one starts from the locator expansion of the resolvent~\cite{aizenman2015random}:
\begin{align}
    G(b, a; E) &= \bra{b} \frac{1}{E-H} \ket{a} \\
    &= \frac{\delta_{ab}}{E-E_a} + \frac{1}{E-E_a} \sum_{p \in \mathrm{P}(a,b)} \prod_{k=1}^{|p|} \frac{g}{E-E_{p_k}}
\end{align}
where $\mathrm{P}(a,b)$ denotes the set of paths from $a$ to $b$. Notice that in our case the labels $a,b,\dots$ will represent Young diagrams, and the graph will be defined by the adjacency matrix $\sum_{\langle D,D' \rangle} \sket{D} \sbra{D'}$ (see $H_{\mathcal{D}}$ in Eq.~\eqref{eq:YD_Hamiltonian}). As customary, one can pass from the random walks $\mathrm{P}(a,b)$ to the self-avoiding walks $\mathrm{SAW}(a,b)$ at the cost of introducing a self energy term:
\begin{multline}
    \label{eq:SAW_expansion}
    G(b, a; E) = G(a,a;E) \\ 
    \times \sum_{p \in \mathrm{SAW}(a,b)} \prod_{k=1}^{|p|} \frac{g}{E - E_{p_k} - \Sigma_{p_k}^{\{p_0,p_1,\dots,p_{k-1}\}}(E)},
\end{multline}
where indeed $\Sigma_a^{\{b,c,\dots\}}(E)$ is the self-energy at site $a$ obtained removing from the lattice the sites $b,c,\dots$. From the \emph{exact} representation of Eq.~\eqref{eq:SAW_expansion} one can in principle obtain also the (many-body) amplitude $\Psi_\alpha(b)$ of the system to be found in configuration $b$, while being in the eigenstate $\alpha$ localized around configuration $a$:
\begin{equation}
    \Psi_\alpha(b) = \frac{1}{\Psi_\alpha(a)} \lim_{E\to E_\alpha} (E-E_\alpha) G(b,a;E).
\end{equation}
Notice that this reduces to $\delta_{ab}$ in the limit $g\to 0$. Finally, performing the approximation of summing only on the \emph{shortest paths} (or \emph{directed polymers}) $\mathrm{SP}(a,b)$ from $a$ to $b$, and thus working to lowest order in $g$, one finds
\begin{align}
    \Psi_\alpha (b) & \approx \sum_{p \in \text{SP}(a,b)} \prod_{k=1}^{|p|} \frac{g}{E_a - E_{p_k}}\\
    \label{eq:FA}
    & = \left(\frac{g}{W}\right)^{d(a,b)} \sum_{p \in \text{SP}(a,b)} \prod_{k=1}^{|p|} \frac{1}{E'_a - E'_{p_k}}.
\end{align}
Above, we have introduced the distance $d(a,b)$, and the rescaled diagonal elements of the Hamiltonian $E'_a := E_a/W$. 

At this point, the criterion for localization is given by the requirement that, with probability $1$ over the disorder realizations, the probability of finding  a particle at distance $O(L)$ from the localization center of the state goes to zero for $L \gg 1$~\cite{pietracaprina2016forward,Scardicchio2017Perturbation}. More formally, defining
\begin{equation}
\label{eq:psi_r}
    \Psi_r := \max_{b:\, d(a,b)=r}|\Psi_\alpha(b)|
\end{equation}
the system is considered to be localized if $Z_r := \frac{1}{r} \ln|\Psi_r|$ satisfies
\begin{equation}
    \label{eq:Zr_loc}
    P\left( Z_r \leq -\frac{1}{\xi} \right)\longrightarrow 1 \qquad \text{for } r\to \infty
\end{equation}
for some finite $\xi>0$. The other way round, if the system is delocalized we expect
\begin{equation}
    \label{eq:Zr_deloc}
    P\left( Z_r\geq -\epsilon \right)\longrightarrow 1 \qquad \text{for } r\to \infty
\end{equation}
for any arbitrarily small $\epsilon > 0$. The critical value of the disorder can be estimated from the average value $\langle Z_{\infty} \rangle = \lim_{r \to \infty} \langle Z_{r} \rangle$ using the condition
\begin{equation}
    \label{eq:Zr_crit}
    \langle Z_{\infty}(W_c) \rangle = -\ln |g| .
\end{equation}
The possibility of passing from the statements in probability, Eqs.~\eqref{eq:Zr_loc}--\eqref{eq:Zr_deloc}, to the one in terms of the average value, Eq.~\eqref{eq:Zr_crit}, is possible because of probability concentration as $r\to \infty$~\cite{pietracaprina2016forward}.

\subsection{Application to the melting process}
\label{sec:FA_results}

\begin{figure}
    \centering
    \includegraphics[width=\columnwidth]{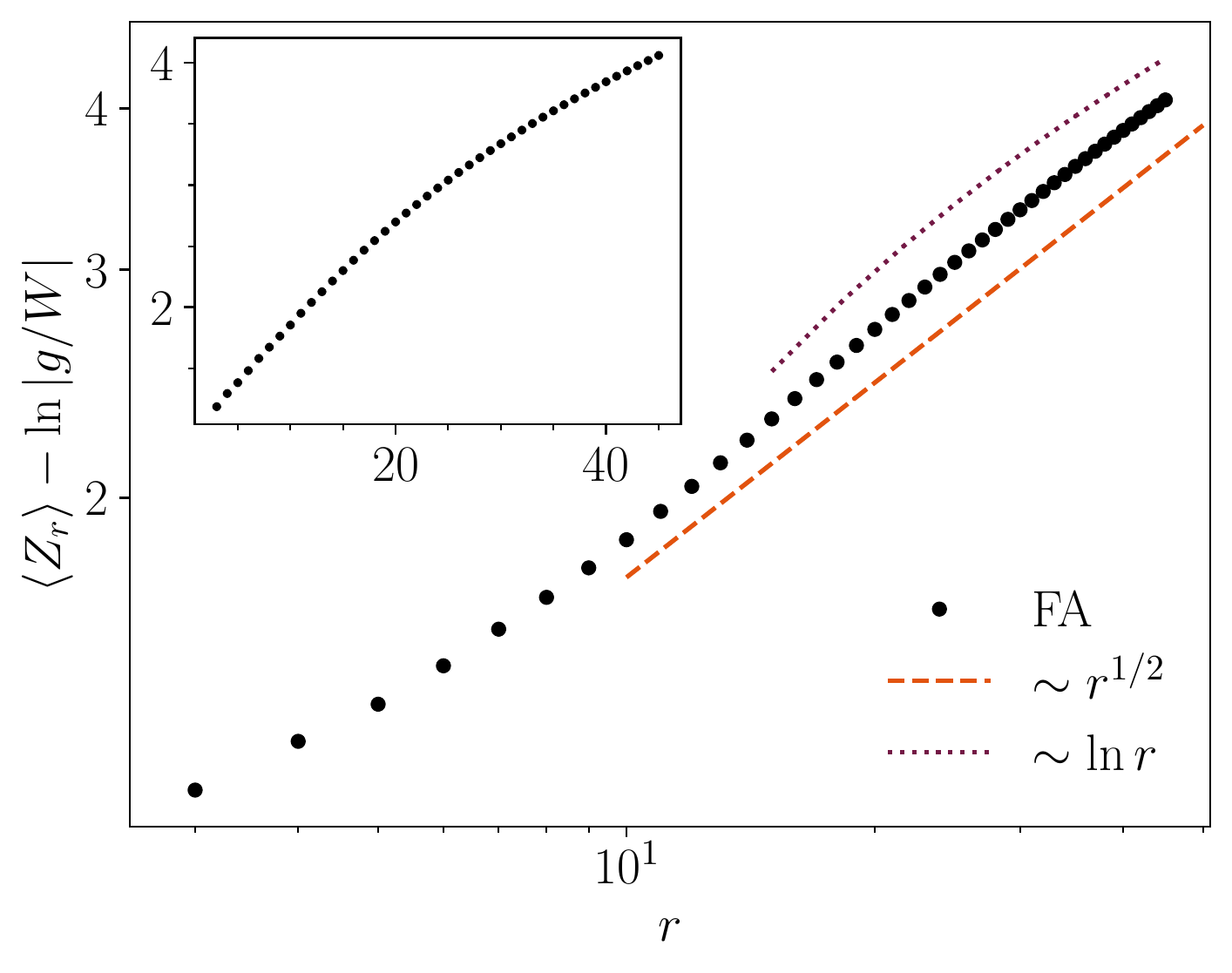}
    \caption{Plot of $\langle Z_r \rangle$ vs $r$ as described in the main text: in the main panel it is shown in log-log scale, while in the inset in linear scale. The dots are the numerical results of the FA up to $r=45$. Their growth should be compared either with a square-root (dashed orange line), or a logarithm (dotted purple line). Fitting a square-root behaviour compares reasonably well with the mean-field-like estimate of the main text: the fit (not shown) gives $Z_r \approx 0.57 r^{0.52}$, while the analytical prediction was $Z_r \approx \pi (2/3)^{3/2} \sqrt{r} \approx 1.71 \sqrt{r}$. The numerical data was averaged over 3000 disorder realizations. The inset contains the same data plotted in linear scale.}
    \label{fig:Zr}
\end{figure}

The numerical results, obtained by using the empty diagram $D = \emptyset$ as starting point (``$a$'' in the formulae above), are reported in Fig.~\ref{fig:Zr}. It is sufficient to plot a value of $W$ only, in virtue of Eq.~\eqref{eq:FA}. As $r$ is increased, $\langle Z_r \rangle$ diverges, being fitted reasonably well both by $\sim \sqrt{r}$ or $\ln r$ (more on this below). This proves the absence of a finite critical value $W_c$, which instead can be present only if $\langle Z_r \rangle$ saturates to a finite constant. 

We now explain why both the square-root and the logarithmic fits are reasonable for the data in Fig.~\ref{fig:Zr} (larger system sizes are needed to discriminate between the two). Starting from the former, one can see that it traces back to the dimension of the Hilbert space as follows. In Eq.~\eqref{eq:FA}, the dominant contribution to the term $(E'_a - E'_{p_k})^{-1}$ is of order $\sim d_{|p_k|}$, being $d_k$ the number of states at distance $k$ from the initial configuration (cf. Sec.~\ref{sec:Young_diagrams}): indeed, one can take the average level spacing to be $\delta E'_{p_k} \approx 2k / d_{|p_k|}$, and take only the dominant (exponential) contribution. The initial configuration being empty, the diagrams at distance $k$ are all made of $k$ blocks, thus they belong to the subspace $\mathcal{H}_{\mathcal{D}_k} \subset \mathcal{H}_\mathcal{D}$. At this point, one can evaluate the product over $k$ in Eq.~\eqref{eq:FA}:
\begin{align}
\label{eq:delta_E}
    \prod_{k=1}^{r} \frac{1}{E'_a - E'_{p_k}} & \sim \prod_{k=1}^{r} d_{k} \sim \exp \left[ \sum_{k=1}^{r} \pi \sqrt{\frac{2 k}{3}} \right] \nonumber \\
    & \sim \exp \left[ \pi \left(\frac{2}{3}\right)^{3/2} r^{3/2} \right] ,
\end{align}
where there was used the Hardy-Ramanujan formula $\dim \mathcal{H}_{\mathcal{D}_k} = d_k \sim \exp \big( \pi \sqrt{2k/3} \big)$, and the asymptotic expansion of the harmonic numbers of order $1/2$~\footnote{The summation can be performed using the Euler–Maclaurin formula. For the present case, it gives $\sum_{k=1}^n \sqrt{k} = \frac{2}{3} n^{3/2} + \frac{\sqrt{n}}{2} + \zeta\left(-\frac{1}{2}\right) + O(n^{-1/2})$, therefore yielding the leading contribution reported in Eq.~\eqref{eq:delta_E}.}. Notice that one can set, according to the convention of Eq.~\eqref{eq:psi_r}, $|p|=r$ and $|p_k|=k$.

The further sum over the SP in Eq.~\eqref{eq:FA} does not alter the behavior of the estimate for large $|p|$, as one can check by giving an upper bound to the number of SP: for example, one can bound it by making all diagrams of size $k$ connected to all diagrams of size $k+1$, for all $k$. In this case, also the number of SP is $\prod_{k=1}^{r} d_{k}$, thus giving the same asymptotic behavior (see also the discussion below).

Putting the pieces together, one gets
\begin{equation}
    Z_r = \frac{1}{r} \ln|\Psi_r| \sim \sqrt{r}.
\end{equation}
This estimate gives a good prediction for $Z_r$, as shown in Fig.~\ref{fig:Zr}, but it relies on the assumption that, at each step of the optimal path, it is feasible to remain as close as possible to the resonant energy. Therefore, we understand that this is a \emph{optimistic} estimate for $Z_r$, yielding a scaling that we can consider to be a sort of upper bound for it.

The assumption of remaining on resonance at each step is not valid for general geometries: it is false, for instance, on the Bethe lattice---which usually constitutes a good approximation of many-body Fock spaces. On the other hand, it is surely valid in the case in which each configuration $D \in \mathcal{D}_r$ is connected to any other configuration $D' \in \mathcal{D}_{r+1}$, in a mean-field-like setting (this same mean-field approximation was used above to bound the number of SP). We argue that the Young lattice of Fig.~\ref{fig:YoungLattice}, i.e.\ the graph obtained by joining two Young diagrams iff they differ by just one square, has indeed properties much closer to the mean-field case rather than to the Bethe lattice. 

Let us consider the number of shortest paths connecting the empty diagram $\emptyset$ to a configuration made of $r$ blocks, call it $D \in \mathcal{D}_r$. For the Bethe lattice, by definition, the number of paths going between any two configurations is one, as there are no loops. On the other hand, considering the mean-field Young lattice in which any configuration in $\mathcal{D}_r$ is connected to any configuration in $\mathcal{D}_{r+1}$, we already showed that the number of shortest paths connecting the empty diagram with any diagram at level $r$ is $\prod_{k=1}^{r} d_{k} \sim \exp (C r^{3/2})$. For the true Young lattice, one can take advantage of the fact that the number of shortest paths leading to a Young diagram $D$ coincides with the so-called dimension $\dim(D)$, computed according to the hook length formula~\cite{Okounkov2003Uses}. Such number $\dim(D)$ corresponds also to the dimension of the representation of the symmetric group identified by the diagram $D$~\cite{Fulton1996Young}. At this point, the typical dimension of a diagram $D$ made of $r$ squares is found to be $\dim(D) \sim \sqrt{r!}$ \cite{Vershik1985AsymMaximal}, so the typical number of SP will scale like $\sqrt{r!}$ as well. Therefore, even if the SP are less than in the mean-field case, they are more than exponentially many in the distance from the starting configuration. In conclusion, one obtains a growth 
\begin{equation}
	\langle Z_r \rangle \sim \ln r.
\end{equation}

The true behaviour of the curve in Fig.~\ref{fig:Zr} will likely be something in between a square root and a logarithm. For the system sizes accessible to present-day computers, and given the slow growth of both curves, it is not possible to discern between the two hypotheses. Nevertheless, for our purposes the results shown are sufficient to claim that there is no finite-disorder localization transition, at least at the lowest order of perturbation theory.

\section{Spectral statistics via exact diagonalization}
\label{sec:spectrum_numerics}

In this Section, we support the conclusions found in perturbation theory by performing an extensive numerical study of the model through exact diagonalization. The numerics was performed by constructing explicitly the Hilbert space of the model, i.e.\ the Young lattice of Fig.~\hyperref[fig:YoungLattice]{\ref{fig:YoungLattice}a}, with ad-hoc methods. An example of the Hamiltonian matrix, truncated to a finite $N$, is shown in Fig.~\hyperref[fig:YoungLattice]{\ref{fig:YoungLattice}b}~\footnote{Another possible way of simulating the system is with the fermionic chain representation. However, we chose not to do so for two reasons. First, the Fock space of a chain of length $L$ at half filling does not contain only the Young diagrams made at most of $L/2$ squares, but also Young diagrams with more squares: consider e.g.\ the state in which $L/2$ fermions are on the left half of the chain, and the right half is empty, that corresponds to a Young diagram made of $(L/2)^2$ squares. Second, and more importantly, the disorder maps to non-local interactions on the chain, which are more difficult to handle.}. The code is made available on GitHub~\cite{GitHub}.

\begin{figure}
    \centering
    \includegraphics[width=0.49\textwidth]{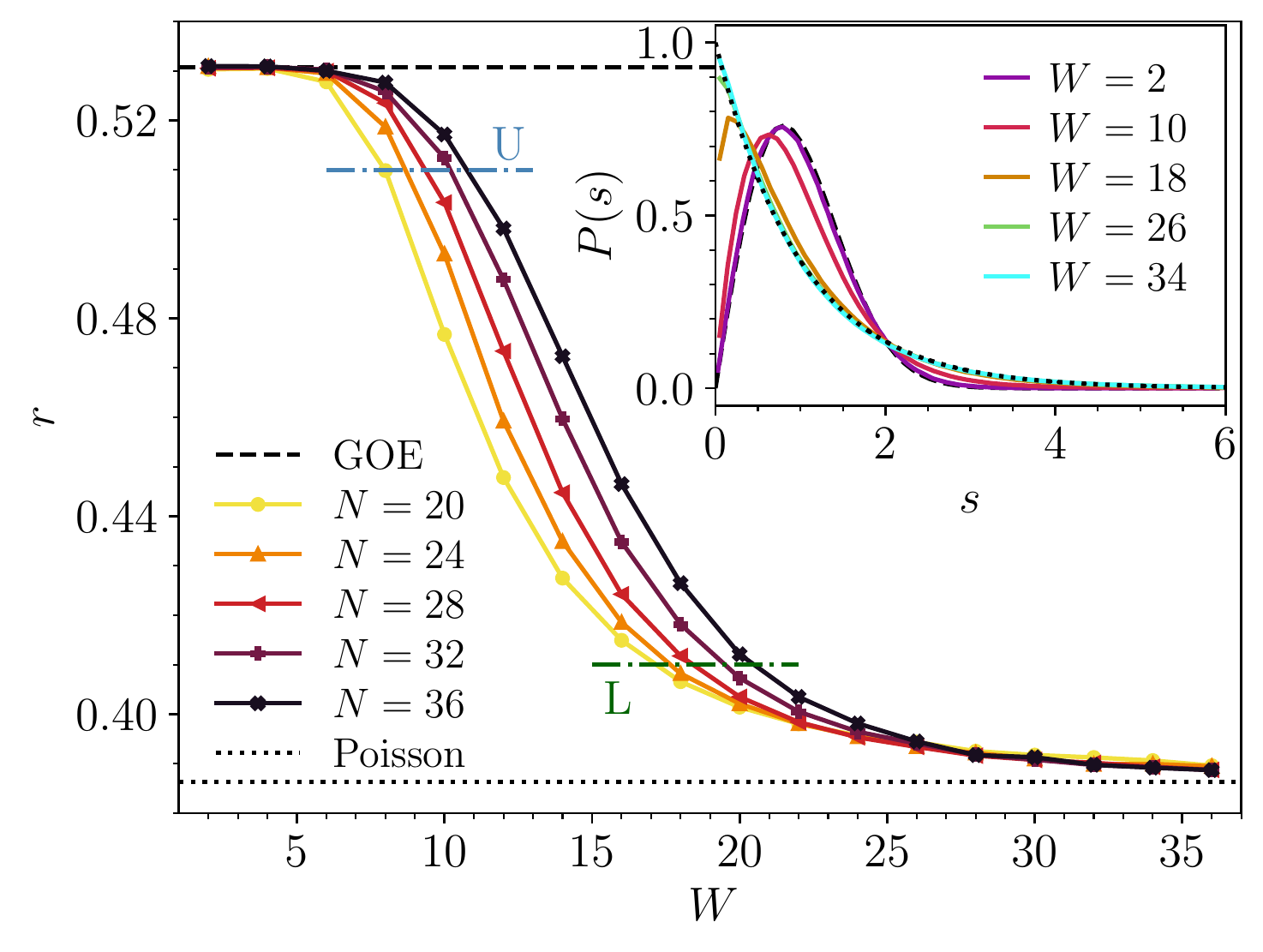}
    \caption{$r$ parameter as a function of the disorder strength $W$ (in units of $g \equiv 1$), and for increasing system sizes. The $r$ value flows from the GOE prediction at small disorder, to the Poisson one at large disorder for any considered system size. However, no real sign of the build-up of a transition is found; rather, the crossover from GOE to Poisson simply seems to shift to larger values of $W$ as the thermodynamic limit is approached. This feature is analyzed by means of the upper (U) and lower (L) cuts, represented by the dashed-dotted lines; see the main text for more details. The number of disorder realizations used ranges from 10000 (smallest system size) to 1700 (largest system size). (Inset) Histogram of the normalized level spacings $s$, for $N=32$ and 3000 disorder realizations. Also here one can see flow from GOE (dashed black line) to Poisson (dotted black line).}
    \label{fig:r}
\end{figure}

\begin{figure}
    \centering
    \includegraphics[width=0.5\textwidth]{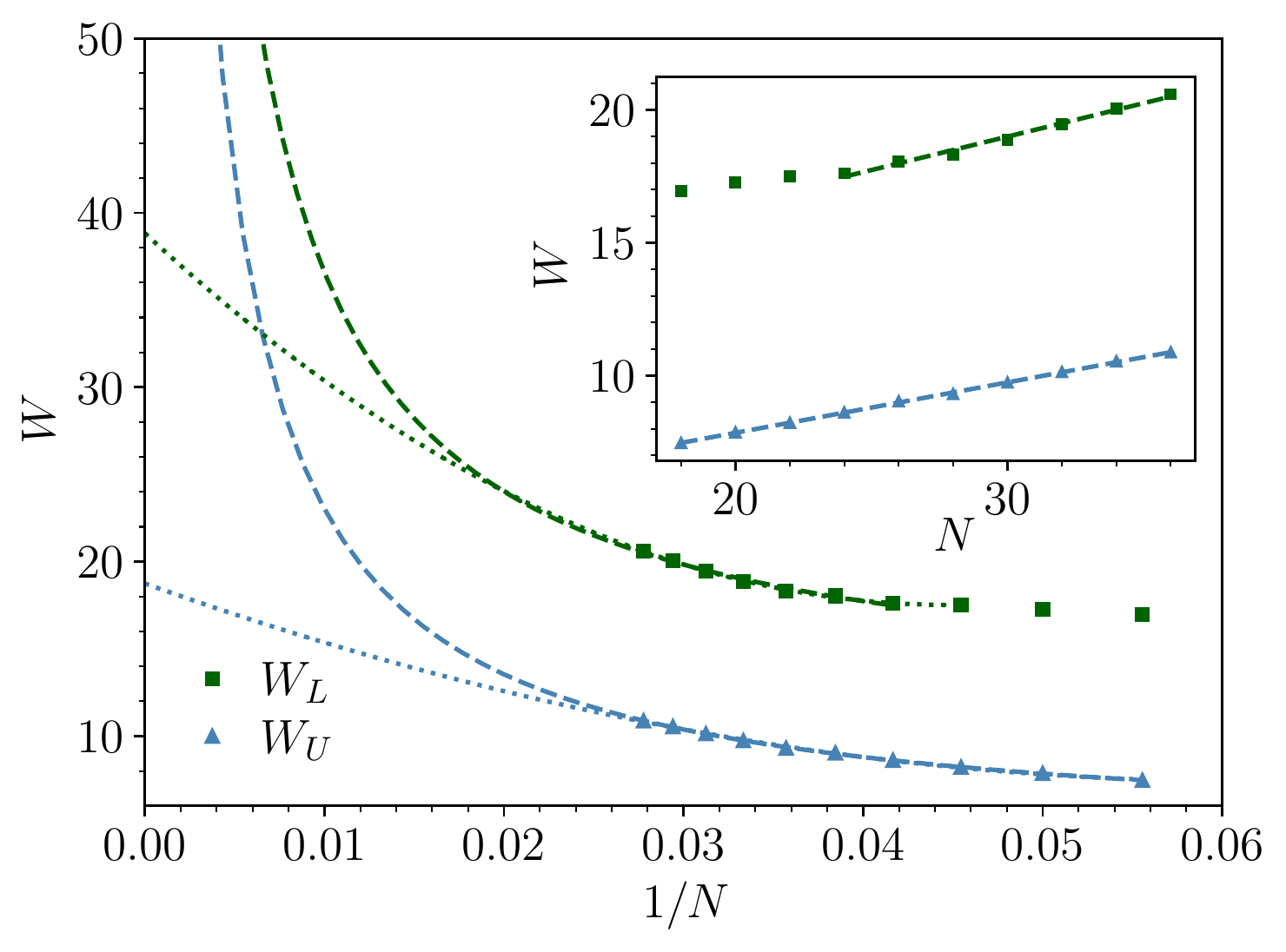}
    \caption{Finite-size scaling analysis of the disorder strengths for which the $r$ parameter becomes smaller than $0.51$ ($W_U$) and $0.41$ ($W_L$). In the inset, it is shown how they seem to diverge linearly with system size, but with two different slopes. In the main panel, two different possible fits are performed: a linear one $W = a + b N$ (dashed line), and one of the form $W = a' + b'/N + c'/N^2$ (dotted line). Further implications are discussed in the main text.}
    \label{fig:cuts}
\end{figure}

To distinguish between the MBL and ETH regimes of a system, one can consider various indicators, each with well-defined, and different behaviors in the two cases. Here, we consider mainly spectral indicators. Let us start from the results for the statistics of the energy levels $E_n$, summarized in Figs.~\ref{fig:r} and \ref{fig:cuts}. In the inset of Fig.~\ref{fig:r} we show how, at finite system size $N$, there is a crossover from Wigner's surmise (viz.\ GOE, at small $W$) to the Poisson gap distribution (at large $W$) for the normalized level spacings $s_n = (E_{n+1}-E_n) / \av{E_{n+1}-E_n}$, taken at the center of the spectrum. To argue that such crossover builds up into a sharp transition in the thermodynamic limit, one may look at the spectral gap ratio parameter 
\begin{equation}
    r = \av{\frac{\min(s_{n+1}, s_n)}{\max(s_{n+1}, s_n)}},
\end{equation}
which needs not be normalized. In the main panel of Fig.~\ref{fig:r}, one can see that the crossover from $r_{\mathrm{GOE}} \simeq 0.5307$ to $r_{\mathrm{Pois}} \simeq 0.3863$ gets slightly steeper as $N$ increases, but it also moves to larger values of $W$. To perform a reliable finite-size scaling analysis, we decided to look at the disorder strengths $W_U$ and $W_L$, for which the $r$ parameter becomes smaller than $0.51$ and $0.41$, respectively~\footnote{The values of 0.51 and 0.41 are of no special importance; any other values near to $r_{\mathrm{GOE}} \simeq 0.5307$ and $r_{\mathrm{Pois}} \simeq 0.3863$ yield the same results.}. Reliable estimates for $W_U$ and $W_L$ were obtained by fitting locally the values of $r(W)$ with a polynomial function, and solving for the intersection. In the inset of Fig.~\ref{fig:cuts}, it is shown how the values found for $W_U$ and $W_L$ seem to diverge linearly with system size, but with two different slopes. In particular, the faster divergence of $W_L$ indicates that no transition is being built up; instead, the crossover from GOE to Poisson seems to become smoother at larger system sizes. Notice that this last fact also prevents one to perform a scaling collapse of the data: it is impossible to accommodate the scalings of both $r>0.51$ and $r<0.41$ with only one function, since the two parts of the curve $r(W)$ are flowing towards larger values of $W$ with different speeds. In addition to the previous observations, both $W_U$ and $W_L$ seem to represent \emph{lower bounds} (see how the curves $r(W)$ change with system size in Fig.~\ref{fig:r}) for the critical disorder strength $W_c$, at which a putative MBL transition may take place: therefore, we believe that such transition does \emph{not} take place at all in the thermodynamic limit, being pushed to infinite disorder strength.

A more refined analysis is shown in the main panel of Fig.~\ref{fig:cuts}. There, we try to extrapolate to $N=\infty$ with two different fits. The dashed line represents the same fit of the inset, i.e.\ a linear one: $W = a + b N$. The dotted line, instead, is a fit of the form $W = a' + b'/N + c'/N^2$, which extrapolates to a finite value at $N=\infty$. Nevertheless, one can notice that the values extrapolated from $W_U$ and $W_L$ are far apart, indicating that either the fitting region is severely pre-asymptotic, or there is no single transition point, but a slow crossover even in the thermodynamic limit. Moreover, one can recognize that, to truly distinguish between the two fitting functions, one would need to go to system sizes $N \gtrsim 60$ (at least for $W_L$, which is the most sensitive to delocalization). Such a system size corresponds to an Hilbert space dimension of more than $\sim 6 \times 10^6$, which is beyond reach for present-day computers and algorithms.

It is interesting to compare our Fig.~\ref{fig:r} with the equivalents of Refs.~\cite{sierant2022stability,Sierant2022Universality}, where instead the data indicates the existence of a transition in the thermodynamic limit. The two plots are substantially different in the scaling as $N \to \infty$. In our case the curves $r(W)$ seem to emanate from a common asymptote as $W \to \infty$, and simply shift towards larger values of $W$ as $N$ is increased. On the contrary, in Refs.~\cite{sierant2022stability,Sierant2022Universality} such curves become steeper already at small system sizes, and in particular the lower part of the curves \emph{moves towards smaller values of $W$}. Therefore, in those works it was possible to analyze another reliable indicator of the MBL transition, namely $W_*$, the point at which the curves for $N$ and $N+1$ intersect. Here, we could not extract a sensible $W_*$ from the data of Fig.~\ref{fig:r} being it ill-defined: the curves $r(W)$ are almost superposed at large $W$.

As a last thing, we remark that all the above results apply to the center of the spectrum, i.e.\ to \emph{generic} states of the model under consideration. However, as stated before, we are interested in the dynamics \emph{starting from a particular state}, i.e.\ the empty Young diagram. Such state has zero expected energy, but for the system under consideration there is no symmetry that forces the spectrum symmetric wrt.\ zero, thus making the corner an infinite-temperature state. We checked explicitly, however, that the corner state on average lies at the center of the spectrum, and that it has a vanishing probability of being very close to the ground state (or the most excited state).

\section{Dynamics}

In the previous Section, we have looked at spectral indicators of ergodicity, and the emerging picture is that there is no \emph{bona fide} MBL phase in the thermodynamic limit for the model under consideration. The absence of a truly localized phase does not immediately imply that, even in the thermodynamic limit, the dynamics of the model should be the same of a standard, ergodic and diffusive system \cite{agarwal2015anomalous,znidaric2016diffusive,luitz2017ergodic}. We will now show, in fact, that the $2d$ quantum Ising model induces on the ``holographic'' chain a peculiar type of dynamics. We will relate the properties on the chain to the ones in $2d$: in particular, the speed of the erosion of the corner will be mapped to the particle current on the chain in Sec.~\ref{sec:transport}. The entanglement entropy arising from a bipartition of the $1d$ chain, instead, will correspond again to the entanglement entropy of a bipartition of the $2d$ model, that we will describe in Sec.~\ref{sec:entanglement}. Finally, in Sec.~\ref{sec:dynamics_numerics} we present the numerical results both for transport and entanglement growth.

\subsection{Transport on the chain}
\label{sec:transport}

As a first step we find, in the $\psi_x$ picture, the number of blocks a Young diagram is composed of. This is done by counting every fermion at distance $x$ to the left from the domain wall, and every hole at distance $x$ to the right, each weighted with the distance from the origin:
\begin{equation}
    \label{eq:N_fermions}
    N = \sum_{x>0} x (1-n_x) + \sum_{x \leq 0} |x| \, n_x,
\end{equation}
where $n_x = \psi_x^\dagger \psi_x^{\phantom{\dagger}}$ is the fermion number at site $x$. Taking into account that the configurations are definitively $n_x \equiv 1$ as $x\to+\infty$, and $n_x \equiv -1$ as $x\to-\infty$, the sum converges. Then, let us take a derivative wrt.\ time in Eq.~\eqref{eq:N_fermions}:
\begin{equation}
    \dot{N}(t) = - \sum_x x \, \dot{n}_x(t).
\end{equation}
Using the fermion number conservation $\dot{n}_x(t) + \partial_x J(x,t)=0$, where $\partial_x$ is the discrete space derivative, we can rewrite the total block number (after an integration by parts) as
\begin{equation}
    N(t) = - \int_0^t dt' \sum_x J(x,t').
\end{equation}
This should be intended as an operator identity. 

In the case of the clean crystal with $h_i\equiv 0$, it can be shown that, in the limit $|x|,t \to \infty$ with $|x/gt|$ held finite~\cite{Balducci2022Localization,Balducci2022Interface}, it holds
\begin{equation}
    \langle n_x(t) \rangle \simeq \frac{1}{2} +
    \begin{cases}
    \frac{1}{\pi} \arcsin \left(\frac{x}{2|g|t}\right)    &\text{if } |x|<2|g|t\\
    \frac{1}{2}\mathrm{sgn}(x)         &\text{if } |x|>2|g|t.
    \end{cases}
\end{equation}
Here, we are using the shorthand notation $\av{A} := \sbra{\emptyset} A \sket{\emptyset}$ for the averages starting from the empty Young diagram initial state. Using this result, in the continuum limit
\begin{equation}
    \langle \dot{N}(t) \rangle \simeq 2 g^2 t,
\end{equation}
and it follows
\begin{equation}
    \label{eq:Nt_clean}
    \langle N(t) \rangle \simeq (gt)^2.
\end{equation}
This power-law scaling can be traced back to the fact that for the ballistic propagation of free fermions
\begin{equation}
    \av{J(x,t)} \simeq
    \begin{cases}
        |g|/2    &\text{if}\ |x|<2|g|t\\
        0       &\text{if}\ |x|>2|g|t,
    \end{cases}
\end{equation}
so
\begin{equation}
    \av{N(t)} \simeq 2g^2 \int_0^t dt'\, t' = (gt)^2.
\end{equation}

Now consider, instead, the case of diffusive motion of the excitations in the fermionic chain. One has $J=-D\partial_x n_x$ for a diffusivity coefficient $D$, so 
\begin{equation}
    \int dx \, \av{J(x,t)} = -D \av{n_{+\infty}(t)} + D \av{n_{-\infty}(t)} = - D
\end{equation}
and
\begin{equation}
    \av{N(t)} \simeq Dt.
\end{equation}

\begin{figure*}
    \centering
    \includegraphics[width=\textwidth]{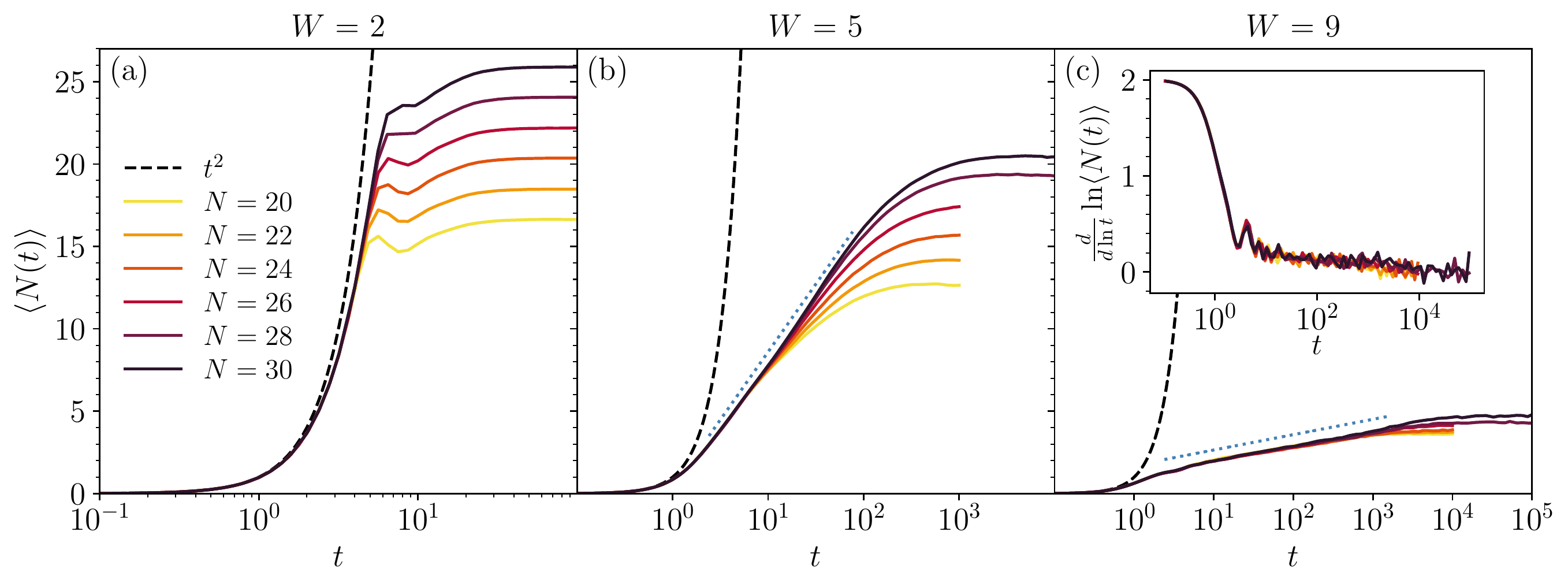}
    \caption{Time evolution generated by the Hamiltonian~\eqref{eq:H_PXP}, starting from the corner state. The average number $\av{N(t)}$ of fermions hops is plotted for various system sizes, and for three values of the disorder strength $W$ (here we set $g \equiv 1$, fixing the energy scale). One can see that at small disorder (i.e.\ $W=2$, panel (a)) the curves do not behave much differently from the prediction for $W=0$, Eq.~\eqref{eq:Nt_clean}, increasing almost as $g^2 t^2$ (dashed line) before saturating. Already at $W=5$ (panel (b)), instead, the growth of $\av{N(t)}$ has been severely hindered, insomuch that it is compatible with a logarithm (dotted blue line): $\av{N(t)} \sim \eta_N \ln t$. Finally, at $W=9$ (panel (c)) the logarithmic behaviour of $\av{N(t)}$ becomes manifest, as shown also by the inset: the logarithmic derivative keeps decreasing towards 0, indicating that $\av{N(t)}$ must be slower than a power law. All the values of $\eta_N$, extracted from fits, are displayed in Fig.~\ref{fig:log_fit}. All the numerical data are averaged over at least 1600 disorder realizations.}
    \label{fig:ev}
\end{figure*}

In a more general setting, the exponent of the growth of $\av{N(t)}$ in the Young blocks is related to the nature of transport for the excitations of the $\psi$ chain ($x(t)$ is the semiclassical trajectory of the excitation):
\begin{equation}
    \av{N(t)}\sim t^{2\beta} 
    \quad \longleftrightarrow \quad
    x \sim t^\beta.
\end{equation}
As just shown above, in the ballistic case $\beta=1$ and in the diffusive case $\beta=1/2$. In Refs.~\cite{li2003anomalous,znidaric2016diffusive,mendoza2019asymmetry} it is discussed at length how the exponent $\beta$ dictates the decay of the correlation functions of the current, of the number $n$, and the non-equilibrium steady state current $J_{\mathrm{ness}}$ in a driven set-up:
\begin{center}
\begin{tabular}{c|c|c|c|c} 
	$\av{N(t)}$ & $x^2$ & $\langle J(0,t)J(0,0)\rangle$ & $\langle n_x(t) n_x(0)\rangle$ & $J_{\mathrm{ness}}$\\ 
	\hline
	$t^{2\beta}$ & $t^{2\beta}$ & $t^{-2+2\beta}$ & $t^{-\beta}$ & $L^{1-\frac{1}{\beta}}$ \\ 
\end{tabular}
\end{center}
The extreme case in which $\beta \to 0$ is expected when entering a MBL phase: $\beta(W)\sim (W_c-W)^\alpha$, although the critical exponent $\alpha$ is currently unknown. In particular, for MBL systems it is possible to show that $N(t)$ saturate to a constant in the long-time limit~\cite{serbyn2013local,huse2014phenomenology,ros2015integrals}. As we will show numerically in Sec.~\ref{sec:dynamics_numerics}, for the model under consideration $\beta \simeq 0$, but the absence of true MBL will manifest in the slow growth $\av{N(t)} \sim \ln (|g|t)$. This implies that the total current decays as $\sim 1/t$, which is indeed an extremely slow decay. We will comment in Sec.~\ref{sec:corner_growth_models} how these features cannot be understood on the basis of simple semiclassical pictures, and instead are due to the quantum nature of the problem.

\subsection{Entanglement growth}
\label{sec:entanglement}

Let us turn now to entanglement spreading. One of the most direct ways of quantifying entanglement growth is to bipartite the system, and consider the entanglement entropy $S_E$ relative to the bipartition. For the setting under consideration, the most natural bipartition is the one that cuts the fermion chain in half through the origin: on the $2d$ lattice, it corresponds to a cut through the vertex of the corner, namely its bisectrix.

In the clean case ($W=0$), being the fermions free it is possible to compute exactly the entanglement growth. The computation was originally carried out in Ref.~\cite{Eisler2009Entanglement} (see also Ref.~\cite{Balducci2022Interface}), and it briefly goes as follows. By definition, $S_E = -\mathrm{Tr}[\rho_A \ln \rho_A]$, $\rho_A$ being the reduced density matrix of subsystem $A$. Both $S$ and $\rho_A$ descend from the correlation matrix $\mathcal{C}_{xy} = \langle \psi^{\dagger}_x \psi_y \rangle$, $x,y \in A$. Therefore, $S_E$ can be computed from the eigenvalues $\lambda_i$ of $\mathcal{C}$ as
\begin{equation}
    S = -\sum_{i=0}^{\infty} \left[ \lambda_i \ln \lambda_i + (1-\lambda_i) \ln (1 - \lambda_i) \right].
\end{equation}
The computation of the eigenvalues turns out to be very complicated in general, and is usually performed numerically. In the continuum limit, however, the situation is simpler as the correlation matrix reduces to the Sine kernel~\cite{Balducci2022Interface}, and a light cone structure emerges, so that $\mathcal{C}$ has non-vanishing elements only inside the light cone. Using such simplifications, one obtains $\lambda_k = 1 / (e^{\epsilon_k}+1)$, with $\epsilon_k(t) = \pm \pi^2 ( k + 1/2 ) / \ln |2|g|\sin(ht)/h|$.

Turning on the disorder ($W \neq 0$), the picture changes significantly. First, as anticipated above the number of particles that hop across the bipartition is severely reduced from $\av{N(t)} \sim (gt)^2$ to $\av{N(t)} \sim \ln (|g|t)$: therefore, one should expect $S_E(t)$ to grow at most like $\sim \ln (|g|t)$ as well. Below, we will show the exact growth of $S_E$ obtained numerically, and comment it in detail.

\subsection{Numerical results}
\label{sec:dynamics_numerics}

Here we summarize the results of a numerical investigation for the dynamics generated by the Hamiltonian~\eqref{eq:H_PXP}. Time evolution was performed through full (for $N \leq 26$) and sparse (for $N \geq 28$) matrix exponentiation with \texttt{SciPy}, having constructed the Hamiltonian matrix incorporating both the hopping and the on-site disorder as in Sec.~\ref{sec:spectrum_numerics}. The code is made available on GitHub~\cite{GitHub}. 

We start by showing in Fig.~\ref{fig:ev} the time evolution for the average number of fermions $\av{N(t)}$ that have hopped. Equivalently, $\av{N(t)}$ is the average number of squares the state is composed of, in the language of Young diagrams. One can see that the growth is ballistic---i.e.\ $\av{N(t)} \sim (gt)^2$---both at short times for all disorder strengths, and at all times for small disorder: for this latter statement, see the case of $W=2$ (in units of $g$) in Fig.~\hyperref[fig:evEE]{\ref{fig:ev}a}. Then, as the disorder is increased slightly, the growth of $\av{N(t)}$ slows down considerably: it acquires a logarithmic behaviour that lasts for three decades already at $W=5$ (Fig.~\hyperref[fig:evEE]{\ref{fig:ev}b}), and for four decades at $W=9$ (Fig.~\hyperref[fig:evEE]{\ref{fig:ev}c}), for the largest system sizes considered, before saturating to a finite-system value~\footnote{Due to the extremely slow dynamics, system sizes larger than $N=30$ were not considered. We believe nonetheless that the system sizes analyzed in this work represent good evidence supporting our claims.}. The growth of $\av{N(t)}$ is more consistent with a logarithm than with a very small power law: in the inset of Fig.~\hyperref[fig:ev]{\ref{fig:ev}c} we show how the logarithmic derivative $d \ln \av{N(t)} / d \ln t$ keeps decreasing towards 0 also for the largest times reached---though some fluctuations are present. Large fluctuations are present also at the level of $\av{N(t)}$: we found the fluctuation of $N(t)$ to be of the same order of magnitude of $\av{N(t)}$ for the strongest disorders considered (i.e.\ $W \gtrsim 8$).

\begin{figure}
    \centering
    \includegraphics[width=\columnwidth]{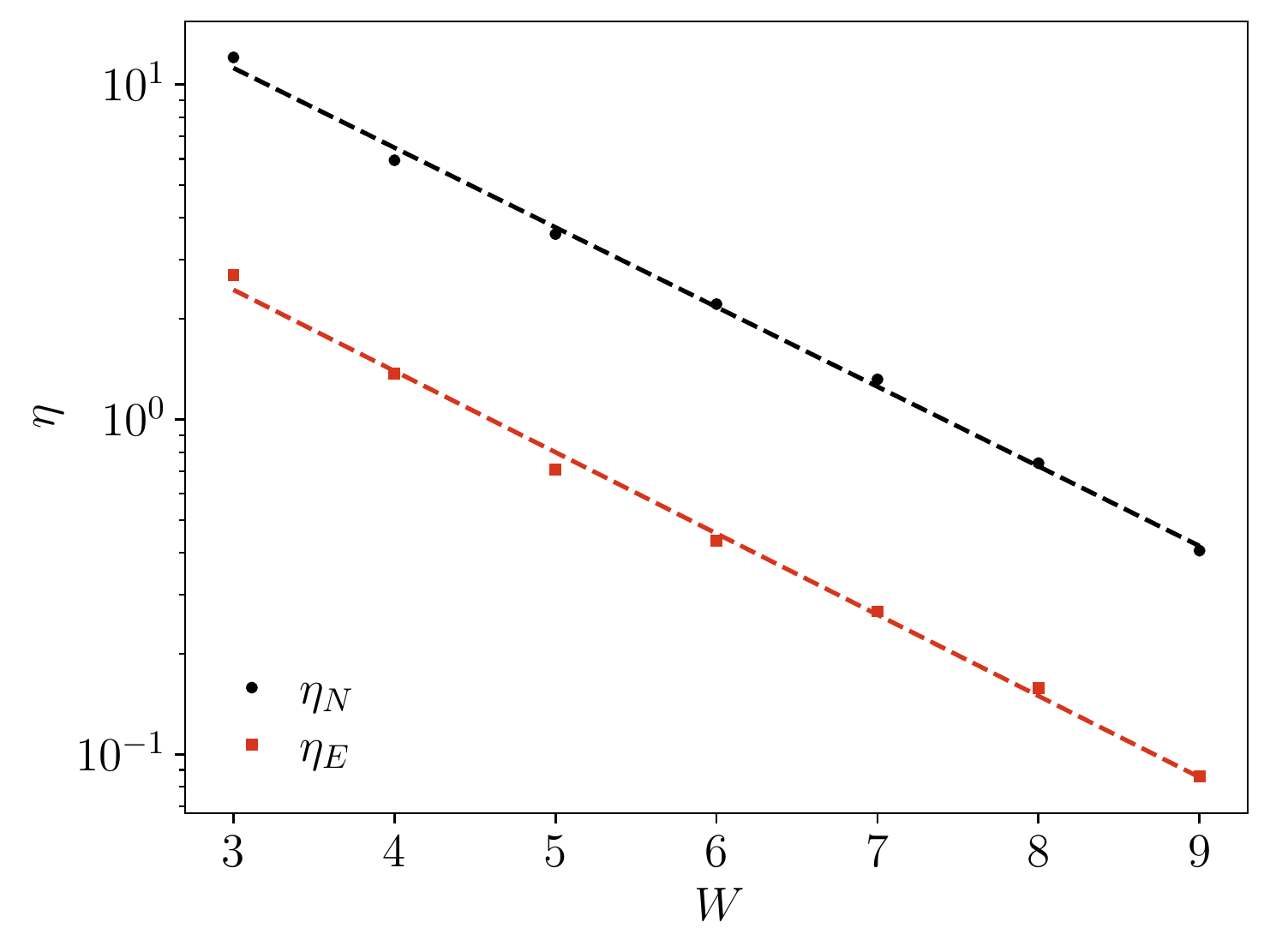}
    \caption{For both the average number $\av{N(t)}$ and the entanglement entropy $S_E(t)$ we performed logarithmic fits $\eta \ln t + c$. Here, we display the dependence of the coefficients $\eta_N$ and $\eta_E$ on the disorder strength $W$. We find both of them compatible with an exponential decay $\eta = \eta_0 e^{-W/W_0}$, with $W_0 \simeq 1.8$ (the dashed lines show the fits).}
    \label{fig:log_fit}
\end{figure}

\begin{figure*}
    \centering
    \includegraphics[width=\textwidth]{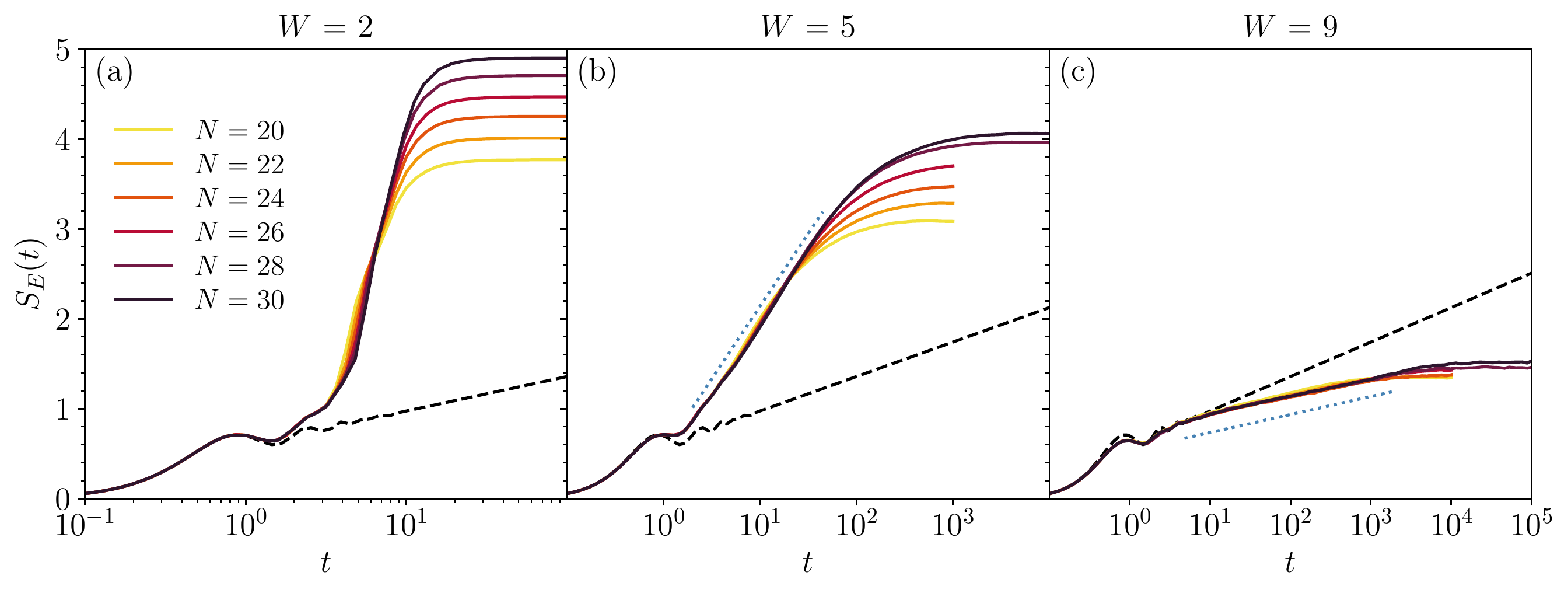}
    \caption{Time evolution of the half-system entanglement entropy $S_E$, for various system sizes, and three different disorder strengths. At small disorder ($W=2$, panel (a)), the growth of entanglement is fast, probably a power law (even if larger system sizes are needed to extract a reliable scaling). Already at moderate disorder ($W=5$, panel (b)), however, the growth of entanglement slows down to a logarithm (the dotted blue line serves as a guide to the eye), being impeded by a logarithmic transport (as described in the main text). When disorder is ramped up ($W=9$, panel (c)), the logarithmic behaviour $S_E(t) \sim \eta_E \ln t$ remains, but with a smaller coefficient $\eta_E$ in front. The coefficients $\eta_E$ extracted from fits are reported in Fig.~\ref{fig:log_fit}. The dashed lines represent the entanglement growth in absence of disorder, i.e.\ $W=0$, which is logarithmic because of integrability. All the numerical data are averaged over at least 1600 disorder realizations.}
    \label{fig:evEE}
\end{figure*}

The remarkable feature of the results of Fig.~\ref{fig:ev} is that, for the same values of the disorder strength $W$, the spectral indicators predict the presence of a \emph{thermal} phase, where it is natural to expect $\av{N(t)} \sim t$ (i.e.\ diffusion), or at most $\av{N(t)} \sim t^{2\beta}$, $\beta < 1/2$ (i.e.\ subdiffusion). We find, instead, a severe impediment to transport, that pushes down $\av{N(t)}$ to a logarithm. In Fig.~\ref{fig:log_fit} we show the results of fits $\av{N(t)} = \eta_N \ln t + c_N$: we find the scaling $\eta_N(W) = \eta_{0,N} e^{- W/W_0}$ with $W_0\simeq 1.8$.

\begin{figure}
    \centering
    \includegraphics[width=\columnwidth]{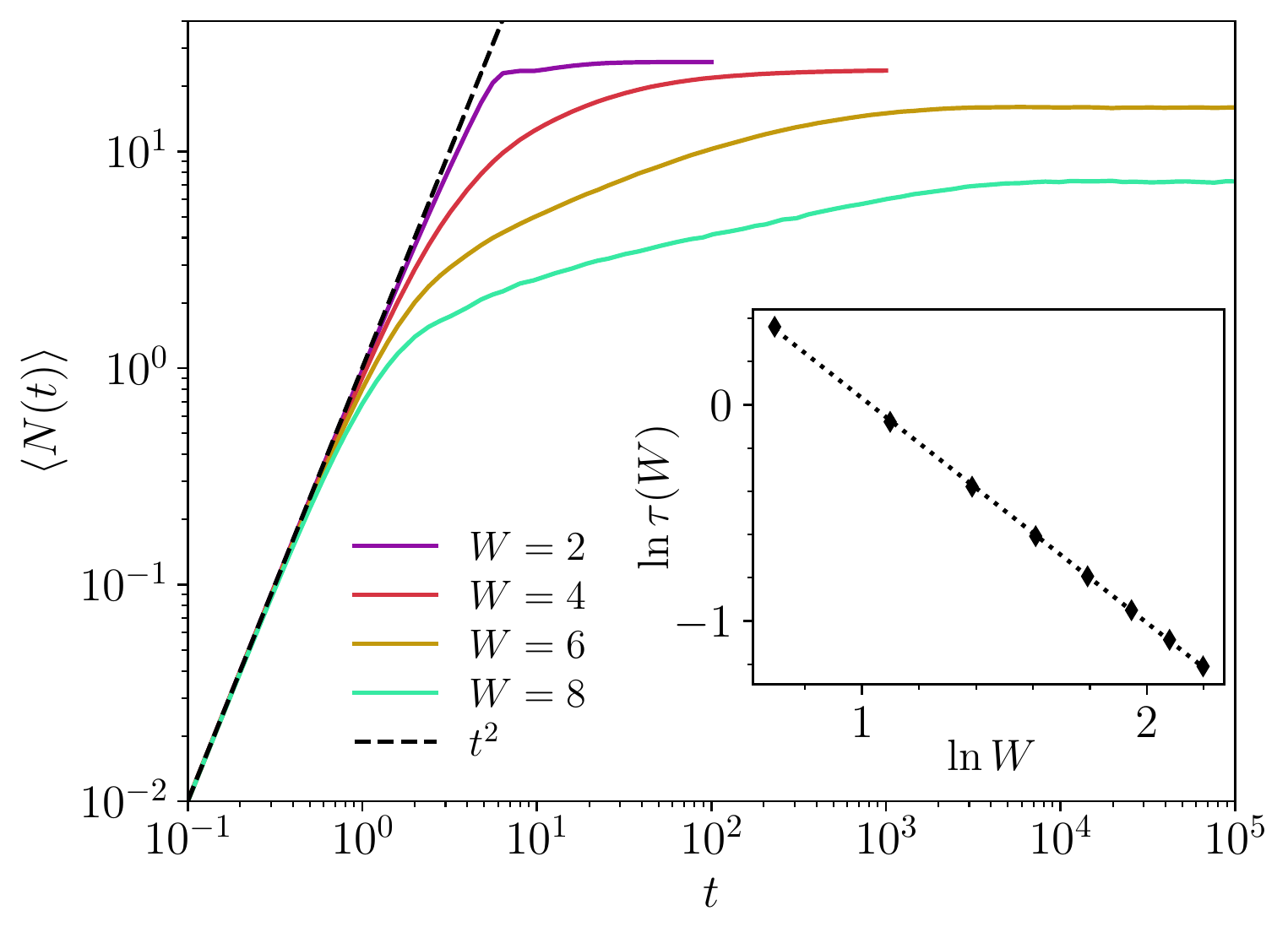}
    \caption{Average number of fermion hops $\av{N(t)}$, for various disorder strengths, with system size $N=30$. The log-log scale makes manifest the behaviour $\av{N(t)} \sim (gt)^2$ at small times ($g \equiv 1$ for numerical purposes), from which $\av{N(t)}$ departs at the time $\tau(W)$. (Inset) Estimates of $\tau(W)$, using a threshold $\varepsilon = 0.05$ (see main text). The fit entails $\tau(W) \sim W^{-\gamma}$ with $\gamma = 1.0$.}
    \label{fig:ev_tau}
\end{figure}

In Fig.~\ref{fig:ev_tau} we analyze instead the behavior of $N(t)$ at earlier times. To this end, we define the timescale $\tau(W)$ that quantifies when the curve $\av{N(t)}$ departs from the ballistic growth $g^2t^2$, e.g.\ when $|\ln \av{N(t)} - 2 \ln(|g|t)|>\varepsilon$ for some fixed threshold value $\varepsilon$. From the results of Fig.~\ref{fig:ev} we expect that $\tau(W)$ suffers of little finite-size effects. Moreover, it is natural to expect $\tau(W)$ to be a decreasing function of $W$, as for strong disorder, the departure from the ballistic motion is supposed to occur sooner. Also, one would guess $\tau(W) \to 0$ for $W \to \infty$, i.e. for every finite disorder strength there is a ballistic regime at small times. Indeed, one can see, in the inset of Fig.~\ref{fig:ev_tau}, that $\tau(W) \sim W^{-1}$.

Let us finally move to the entanglement growth. In Fig.~\ref{fig:evEE} we consider the entanglement entropy, relative to the bipartition along the bisectrix of the corner (and, consequently, that cuts the fermionic chain at the origin). Several comments are in order. First, despite the ballistic spreading of \emph{particles}, at $W=0$ the entanglement growth is only \emph{logarithmic} in time (dashed black line in Fig.~\ref{fig:evEE}), because of integrability. To see this, one can employ the so-called \emph{quasiparticle picture}, or the conformal field theory description in the continuum~\cite{Calabrese2007Entanglement}: in both cases, the slow growth of entanglement is traced back to excluded volume effects among the fermions. It should not worry, then, that $S_E(t)$ \emph{grows faster if} $W>0$, but small: indeed, a small amount of disorder helps the system in thermalizing, and the entanglement entropy raises linearly in time, essentially because of chaos propagation~\cite{Nahum2017Quantum,Nahum2018Operator,Chan2018Solution}.

Second, in Fig.~\ref{fig:log_fit} we show the results of fits $S_E(t) = \eta_E \ln t + c_E$, as was done for the number growth. We find the scaling $\eta_E(W) = \eta_{0,E} e^{-W/W_0}$ with the same $W_0 \simeq 1.8$. Such agreement does not come unexpected: if transport is blocked, and no long-range dephasing interactions are present (contrary to the l-bit model of MBL~\cite{serbyn2013local,huse2014phenomenology,ros2015integrals}), then entanglement cannot spread beyond the melted part of the corner. Indeed, for each particle that hops across the origin, the entanglement entropy between the left and right halves of the chain increases fast, well before the next hop, because of non-local interactions entailed by the disordered potential. But such non-local interactions act only on the melted part of the corner, and thus particle spreading functions as a bottleneck for the entanglement growth.

\section{Discussion}
\label{sec:discussion}

In this Section, we take the chance to describe the limits of validity of the approximations used (Sec.~\ref{sec:approximations}), and to draw a comparison with \emph{classical} corner growth models, that have been extensively studied in the literature (Sec.~\ref{sec:corner_growth_models}). Finally we present some concluding considerations in Sec.~\ref{sec:conclusions}.

\subsection{Limits of validity of the approximations}
\label{sec:approximations}

So far, we have been discussing the dynamics of melting of a $2d$ disordered quantum crystal, by modeling it through the strong-coupling limit of the $2d$ quantum Ising model, in presence of a random longitudinal field. However, the coupling $J$ was \emph{effectively taken to be infinite} or, equivalently, the $O(1/J)$ corrections were considered always negligible. On the other hand, in Refs.~\cite{Balducci2022Localization,Balducci2022Interface} the $O(1/J)$ corrections were studied in depth, showing that they lead to interesting phenomena as Stark MBL. The rationale behind the choice of neglecting the corrections in this work was the following: while in the clean system ($W=0$) the introduction of interactions leads to integrability breaking, for $W \neq 0$ it would lead to just minor \emph{quantitative} modifications in the dynamical behaviour. While we refer to Refs.~\cite{Balducci2022Localization,Balducci2022Interface} for the precise form of such $O(1/J)$ corrections, here we just remark that they are (parametrically small) four-body interactions on the fermionic chain: therefore, they become negligible in comparison with the strong, non-local interactions arising from the disorder. In particular, the sum appearing in Eq.~\eqref{eq:E_D} (and therefore in Eq.~\eqref{eq:interactions}) makes the disordered interactions of order $\sqrt{N} W$, when acting on Young diagrams of size $N$. Consequently, they become stronger as time passes by, and the Hilbert space of larger Young diagrams is explored, making the $O(1/J)$ corrections even less relevant.

Of course, we expect the picture presented to break down at small values of $J$: there, also the mapping to fermions ceases to be valid, since it becomes possible for any $2d$ spin to flip with non-vanishing probability, and the interface is no more well defined. How the dynamics changes in such limit is however a very interesting question, that we hope may be the object of future studies.

\subsection{Comparison with classical corner growth models}
\label{sec:corner_growth_models}

The slow growth of the average number of squares in the Young diagrams $\av{N(t)}$, observed in the quantum dynamics (Sec.~\ref{sec:dynamics_numerics}), turns out to be anomalous also from the perspective of similar, classical models. Indeed, using the same mapping to a chain detailed in Sec.~\ref{sec:fermions}, one can describe a classical corner growth model in terms of simple exclusion processes on the line~\cite{Balazs2006Cube,Ferrari2010Random}. In the absence of disorder, it is natural to associate the quantum process to the totally anti-symmetric simple exclusion process (TASEP), that turns out to have ballistic dynamics~\cite{Derrida1998Exactly}, but a different limiting shape for the eroded part~\cite{Balducci2022Interface}. When disorder is added instead, one might hope to reproduce the quantum dynamics with an exclusion process in which particles are subject to a strongly inhomogeneous waiting time before moving, according to some probability distribution. It turns out that, even when a fat-tailed probability distribution for the waiting times is chosen  (this also makes the process non-Markovian), the growth of the eroded corner is power-law~\cite{Khoromskaia2014Dynamics,Jose2020Bidirectional}, never attaining a logarithmic behavior as the one observed in the quantum regime. In particular, a logarithmic growth can be obtained only if the waiting-time distribution behaves like $p(\tau) \sim \tau^{-1}$ for large $\tau$, i.e.\ it is non-normalizable. This is an indication of the purely quantum nature of the problem we have discussed, that cannot be reproduced by classical means.

Another interesting question is about the comparison of the average limiting shapes, between classical and quantum melting processes. The clean case was already studied in Refs.~\cite{Balducci2022Localization,Balducci2022Interface}, to which we refer for further details. The disordered case is more intriguing, and difficult to analyze: we plan to discuss this issue in a future work.

\subsection{Final considerations}
\label{sec:conclusions}

In this work we addressed the spectral and dynamical properties of a disordered two-dimensional quantum crystal. In particular, we studied the quantum Ising model on a square lattice, and studied the melting of an infinite, corner-shaped interface. While the same problem turned out to display ergodicity breaking in absence of disorder \cite{Balducci2022Localization,Balducci2022Interface}, in this work we presented both analytical and numerical evidence supporting the {\it absence} of a many-body localized phase when disorder is added. We established, through an analytical argument based on the forward approximation, and numerical results for spectral properties, that the model is ergodic for any finite $W$ in the thermodynamic limit. However, we also showed that the dynamics turns out to be extremely slow: we found, through an extensive numerical analysis, that the growth of the average number of melted squares, $\av{N(t)}$, passes from ballistic to logarithmic in time already for small disorder, and we characterized the crossover between these two regimes with various indicators. Also the entanglement entropy $S_E(t)$ shows a similar behavior, that traces back to the growth of $\av{N(t)}$. 

While the results showed in this work support the common belief that MBL does not survive in dimensions higher than one, we presented strong evidence for the onset of slow dynamics, namely slower than subdiffusive. Surprisingly, such behavior is already present at small disorder strength, when the system, at finite size, is fully ergodic according to the spectral indicators. An explanation of this slow, logarithmic dynamics in terms of the phenomenology of avalanches~\cite{de2017many,de2017stability,Thiery2018many} would be desirable, but we did not address it in this first work.

\acknowledgements

F.B.\ thanks L.\ Capizzi, A.\ Santini and V.\ Vitale for discussion.


\bibliography{references}

\end{document}